\documentclass[aps,prx,reprint,longbibliography]{revtex4-2}

\pdfoutput=1 %% Uncomment for arxiv Submission
\usepackage{amsmath}
\usepackage{amssymb}
\usepackage{mathrsfs}
\usepackage{physics}
\usepackage{hyperref}
\usepackage[capitalize]{cleveref}

\usepackage[usenames]{xcolor}
\usepackage{graphicx}

\newcommand{\widesim}[2][2]{
  \mathrel{\underset{#2}{\scalebox{#1}[1]{$\sim$}}}
}

\DeclareMathOperator{\sgn}{sgn}
\DeclareMathOperator{\arcoth}{arcoth}

\begin{document}

\setlength{\abovedisplayskip}{6pt}
\setlength{\belowdisplayskip}{6pt}

\title{Strange metal in the doped Hubbard model via percolation}

\author{Andrew A. Allocca} \email{aallocca@lsu.edu}
\affiliation{Department of Physics and Astronomy and Center for Computation and Technology, Louisiana State University, Baton Rouge, Louisiana 70803, USA}

\begin{abstract}
    Many strongly correlated systems exhibit strange metallic behavior in certain parameter regimes characterized by anomalous transport properties that are irreconcilable with a Fermi-liquid-like description in terms of quasiparticles. 
    The Hubbard model is a standard theoretical starting point to examine the properties of such systems and also exhibits non-Fermi-liquid behavior in simulations. 
    Here we analytically study the two-dimensional hole-doped Hubbard model in the large $U$ limit, first identifying a doping-induced percolation transition in the low-energy sector occurring at a definite critical doping $p_c$ depending on lattice structure. 
    Using the critical properties near this transition we rewrite the Hubbard Hamiltonian and motivate a low-energy lattice-independent large-$N$ model with distinct non-Fermi-liquid properties. 
    We show that this model has linear-in-$T$ resistivity with doping-dependent slope maximized at $p=p_c$ and power-law optical conductivity $\sim \abs{\omega}^{-2/3}$.
    Though the parameters used in developing this theory mean it cannot be directly applied to the cuprate superconductors, we nevertheless reproduce several important phenomena observed in their strange metal phase, and also predict these same qualitative behaviors to manifest in other lattices near concrete hole dopings. 
\end{abstract}

\maketitle

\section{Introduction}
\vspace{-12pt}
Strange metals, first identified in the hole-doped cuprate superconductors~\cite{Cava1987, Gurvitch1987}, are principally characterized by a dc-resistivity with anomalous, often linear, temperature dependence $\rho\sim T$ at low temperatures, starkly different from the quadratic behavior in normal metals described by Fermi liquid theory.
Though electron-phonon scattering is known to produce linear-in-$T$ resistivity in normal metals above the Debye or Bloch-Gr\"uneisen temperatures~\cite{Peierls1934, Peierls1996}, in strange metals this behavior persists well below this threshold, down to the lowest measurable temperatures.
In the cuprates the behavior persists from the melting of low-temperature ordered phases up to very high temperatures with no apparent change in slope at the Debye temperature near optimal doping, raising serious doubt about the relevance of phonon scattering for the strange behavior of materials even in the regime where it should contribute~\cite{Hartnoll2015, Mousatov2021, Hartnoll2022}.
This linear behavior is most prominent near optimal doping or the critical value $p_c$ signaling the collapse of the pseudogap phase~\cite{Loram2001,Tallon2001,Tallon2020,Proust2019}, with a gradual crossover from $\rho \sim T$ around these values to more standard Fermi-liquid-like $\rho \sim T^2$ far enough into the overdoped regime~\cite{Manako1992, Nakamae2003, Cooper2009}. 
Studies of other properties find additional anomalous, non-Fermi-liquid-like behaviors.
The optical conductivity has apparent power-law scaling $\sigma(\omega)\sim \abs{\omega}^{-\gamma}$ over a finite frequency range, with approximate exponent $\gamma\approx 2/3$ near $p_c$ varying slightly between samples and with doping~\cite{Cooper1990, Schlesinger1990, ElAzrak1994, Baraduc1996, Marel2003, Hwang2007}, though non-power-law forms of this quantity may also explain these observations~\cite{Michon2023}. 
Additionally, the Hall angle is found to obey $\cot\Theta_H \sim A T^2 + B$ across a wide range of doping levels~\cite{Chien1991, Carrington1993, Segawa2004}.

In addition to this phenomenology, there is growing experimental evidence for a temperature-independent phenomenon in the cuprates occurring at the critical hole doping $p_c$. 
Transport measurements in LSCO subjected to magnetic fields strong enough to suppress superconductivity find a broad range of dopings at low temperatures supporting $T$-linear resistivity~\cite{Cooper2009}, with features in the doping dependence of the resistivity at $p_c\approx 0.185$;
the coefficients of $T$ and $T^2$ terms fitting the measured behavior of $\rho$ have distinct features at $p_c$, with the linear coefficient achieving its maximum value. 
As noted in Ref.~\cite{Cooper2009}, this phenomenology contrasts with the behavior expected from a quantum phase transition, which would produce this behavior in only a narrow fan emerging from the quantum critical point, not a broad region at low temperature. 
A more recent ARPES study of Bi2212~\cite{Chen2019} identified a $T$-independent boundary in the temperature-doping phase diagram at $p_c\sim0.19$ across which the spectral function near the Brillouin zone boundary abruptly changes from incoherent below $p_c$ to coherent with an identifiable quasiparticle peak above $p_c$. 

Evidence of non-Fermi-liquid behavior is also found in studies of the Hubbard model on a two-dimensional square lattice, believed to model essential features of the CuO${}_2$ planes of the cuprates as well as being the prototype for more general systems of strongly correlated electrons~\cite{Dagotto1994, Imada1998}. 
Numerical analyses identify marginal Fermi-liquid~\cite{Varma1989} behavior of the electronic self-energy in hole-doped systems at nonzero temperature~\cite{Kakehashi2005, Vidhyadhiraja2009}, and show that linear-in-$T$ resistivity persists down to the lowest temperatures these methods can reach~\cite{Huang2019, Wu2022}, connecting with studies finding $T$-linear behavior at much high temperatures~\cite{Perepelitsky2016}. 
Simulations of the Hubbard model in systems of cold atoms also extract diffusion and transport properties that are consistent with linear-in-$T$ charge transport~\cite{Xu2019, Brown2019}.
Thus, the Hubbard model is itself sufficient to produce the defining strange metal phenomenology---$\rho\sim T$ resulting from some effect other than electron-phonon scattering---and phenomena in real materials such as structural transitions, disorder, etc. are evidently not necessary to realize this behavior.
It is important to note, however, that in all of these studies the lowest accessible temperatures correspond to parameter regimes well above the strict definition of the strange metal phase of the cuprates, i.e. the Debye temperature.
While the Hubbard model can produce apparently strange-metallic phenomenology, i.e. $\rho\sim T$ unrelated to phonon scattering, one cannot say that it models the actual strange metallic behavior of the cuprates directly, i.e. observations of $\rho\sim T$ below the Debye temperature, for which phonon scattering can be unambiguously ruled out as a possible mechanism.

The $T$-linear nature of the resistivity and the other unusual properties in strange metals appear to be inconsistent with a description of these system in terms of quasiparticles.
Current is dissipated on a timescale $\tau \sim \hbar/k_B T$ depending only on temperature~\cite{Legros2019, Hartnoll2022}, suggesting that at least the charge-carrying sector of these systems contain no other characteristic energy scale, such as a quasiparticle mass or Fermi energy.  
Consequently, in recent years there has been a great amount of theoretical work aiming to better understand strongly-interacting systems without quasiparticles, such as the Sachdev-Ye-Kitaev (SYK) model and its large-$N$ relatives~\cite{Sachdev1993, Parcollet1998, Parcollet1999, Kitaev2015a, Kitaev2015b, Sachdev2015, Song2017, Esterlis2019, Chowdhury2022, Sachdev2023}, or models with holographic dualities to gravitational theories~\cite{Hartnoll2018}.
Other approaches examine the loss of quasiparticles near a Fermi surface via coupling to critical bosons~\cite{Coleman2001, Florens2004, Senthil2008a, Senthil2008b, Lee2009, Metlitski2010, Else2021, Shi2022}, which can also be accomplished with large-$N$ models~\cite{Esterlis2021, Guo2022, Shi2023}. 
Still other works invoke the intrinsically nonquasiparticle concept of unparticles~\cite{Georgi2007, Phillips2013, Karch2016, Limtragool2016, Leong2017}, or take a model-agnostic hydrodynamic approach to study systems with anomalous transport properties~\cite{Andreev2011, Hartnoll2015, Lucas2015, Lucas2016, Lucas2017}. 
These approaches have their own strengths and weaknesses.
Large-$N$ models can be solved exactly but the large number of flavors they invoke have no clear microscopic interpretation.
Theories with critical bosons use uncontroversial, well-understood properties of quantum critical points, but there is a history of difficulty in actually identifying the nature of the critical doping $p_c$ in the cuprates---as noted above, some observations are at odds with identifying this as a standard quantum critical point. 

Drawing on these experimental, numerical, and theoretical insights, here we develop a method to analyze the large-$U$ Hubbard model that yields an effective theory exhibiting strange metallic properties. 
The foundation of this method is to identify and leverage the structure of a classical percolation transition~\cite{Aharony2003} occurring in generic many-electron states permitting only singly-occupied sites, the result of strong on-site repulsion. 
In these states, electrons of one spin species bound clusters of sites that contain all electrons of the opposite spin, and if the electron positions are uncorrelated beyond the single-occupation condition we can analyze the properties of these clusters of sites in terms of just the electron density---a straightforward calculation relates the percolation threshold of these clusters to a critical hole doping $p_c$ independent of temperature.
For $p\sim p_c$ the properties of these clusters are thus dictated by a critical theory, and by rewriting the Hubbard Hamiltonian in terms of these clusters the system's dynamics reflect aspects of this criticality. 
The Hubbard Hamiltonian rewritten in terms of these critical clusters can be related to an effective large-$N$ model, where the $N$ ``flavors'' of the model are ultimately related to the large number shapes that large clusters may take.
We calculate the transport properties of this model and find linear-in-$T$ dc resistivity, with maximal slope at $p=p_c$, and a power-law contribution to the optical conductivity $\sigma(\omega) \sim \abs{\omega}^{-\gamma}$ with $\gamma\approx 2/3$ applicable for a range of frequencies.

While these properties appear consistent with the phenomenology of the cuprates including the value of $p_c$---we find $p_c^{(\mathrm{sqr})} \approx 0.1854$ for the square lattice---we must be careful in considering how this theory applies to these materials.
The infinite-$U$ limit sets the exchange energy $J\sim t^2/U$ to zero, so the theory only applies to temperatures above this scale, which itself is well above the Debye temperature of real cuprates. 
Therefore, though our analysis yields linear-in-$T$ resistivity unrelated to electron-phonon scattering and several other notable features of the cuprates, it fundamentally does not, in its current form, model the true low-temperature cuprate strange metal phase.
The infinite-$U$ limit would need to be relaxed to extend the model's applicability towards physically relevant regimes.

In other ways this analysis applies in more general circumstances than the cuprates provide.
The cuprates are an important point of comparison for the theory in that they realize physics of the Hubbard model, but only on the square lattice.
Here we obtain the same effective theory for \emph{any} lattice that exhibits a percolation transition at a non-trivial occupation probability.
Therefore, this analysis suggests that the properties listed above should manifest in other lattices at specific doping levels.
For the hexagonal and kagome lattices, for example, we calculate critical hole dopings $p_c^{(\mathrm{hex})} \approx 0.394$ and $p_c^{(\mathrm{kag})} = 1-4\sin(\pi/18) \approx 0.305$ where the same linear-in-$T$ resistivity and power-law optical conductivity arise.

The remainder of this paper is organized as follows.
In \cref{sec:percolation} we give a brief introduction to relevant aspects of site percolation and clusters, which will be used throughout the rest of the analysis.
We start our analysis of the Hubbard model in \cref{sec:clusterbasis}, introducing the cluster basis and cluster operators, and rewriting the Hubbard Hamiltonian in terms of these degrees of freedom. 
In \cref{sec:approximations} we discuss approximations to this form of the Hamiltonian, which motivates the large-$N$ model action we write and begin to analyze in \cref{sec:modelaction}. 
In \cref{sec:transport} we calculate the anomalous transport properties of the model, and finally in \cref{sec:discussion} we discuss these results and the outlook for further developments. 

\section{Two-Dimensional Site Percolation} \label{sec:percolation}
\vspace{-12pt}

Since it is central to our construction and establishes notation used throughout, we begin with a brief summary of relevant aspects of percolation theory, as given in Refs.~\cite{Stauffer1979, Aharony2003}. 
In a site percolation model, the sites of an infinite lattice are randomly and independently occupied with some probability $P$, and the average properties of clusters of occupied sites are then analyzed as functions of $P$.
Two occupied sites belong to the same cluster if they are nearest-neighbors, and the number of sites comprising a cluster, which we call its size, is denoted with $s$.
Some quantities of interest are the size of the largest cluster, the average cluster size, and the distribution of cluster sizes. 
Importantly, a phase transition occurs in this system as the occupation probability passes through a critical value $P_c$.
For $P<P_c$ the system has only finite clusters---average $s$ is finite and the largest cluster has size $s_\xi$. 
As $P\to P_c$ from below $s_\xi$ diverges, and for $P>P_c$ there is single infinite cluster which encompasses an ever larger portion of all sites in the system for increasing $P$.
At $P=P_c$ the system has clusters of every finite size, no characteristic scale, and can be described by a conformal field theory. 
The value of this critical occupation probability is highly dependent on geometry, with the values for notable lattices given in \cref{tab:Pcvals}. 

\begin{table}[!b]
    \centering
    \begin{tabular}{cc}
    \hline\hline
        Lattice & $P_c$~\cite{Suding1999} \\
        \hline
        Square & $0.5927$ \\
        Hexagonal & $0.6970$ \\
        Kagome & $1-2\sin(\pi/18) \approx 0.6527$ \\
        Triangular & 1/2 \\ 
        \hline\hline
    \end{tabular}
    \caption{Values of the critical occupation probabilities for several notable lattices.}
    \label{tab:Pcvals}
\end{table}

Near to the critical point the average properties and distribution of finite clusters obey a scaling theory controlled by $\abs{P-P_c}$ and characterized by critical exponents, independent of lattice geometry.
For $P<P_c$ the average number of sites in the largest finite cluster is $s_\xi\propto \abs{P-P_c}^{-1/\sigma_P}$ and the average linear size of this cluster is the correlation length $\xi_P \propto \abs{P-P_c}^{-\nu_P}$, with critical exponents $\sigma_P = 36/91$ and $\nu_P = 4/3$ in two dimensions~
\footnote{We use the subscript $P$ to unambiguously specify quantities associated with the percolation transition.}. 
These quantities are related through $s_\xi \propto \xi_P^{d_f}$, where $d_f = 1/(\nu_P\sigma_P) = 91/48$ is a fractal dimension characterizing this scaling on average for large $s$.
Motivated by this, for large enough $s$ we can define an average cluster radius $R$ through $s \propto R^{d_f}$. 
For large $s$ the average density of clusters of size $s$ is $n_s \propto s^{-\tau_P} f(s/s_\xi)$ with critical exponent $\tau_P = 187/91$, where the function $f$ is constant for small argument and rapidly decays for argument greater than 1.
The total density of clusters (number of clusters per lattice site) is the sum of $n_s$ over all $s$,
\begin{equation} \label{eq:clusterdensity}
    \scalebox{0.92}{$n_C(P) = \sum_s n_s \sim A_0 + A_1 \abs{P-P_c} + A_2 \abs{P-P_c}^{2-\alpha_P},$}
\end{equation}
written for $P$ near $P_c$ in terms of constants $A_0, A_1, A_2$ and critical exponent $\alpha_P = -2/3$ in two dimensions~\cite{Aharony2003}.

These scaling relationships only apply for large enough clusters. 
Small clusters contribute analytic terms for some quantities, but are largely irrelevant for critical properties near the transition. 
We use $s_0$ to denote the cluster size above which these scaling properties hold, and $R_0 \gg a$ for the corresponding radius, where $a$ is the lattice constant. 
\vspace{-12pt}
\section{Rewriting the Hubbard Hamiltonian} \label{sec:clusterbasis}
\vspace{-12pt}
Turning now to an electronic system, we write the Hubbard Hamiltonian with isotropic nearest neighbor hopping as
\begin{equation} \label{eq:HubbardH}
\begin{gathered}
    H = H_U + H_\uparrow + H_\downarrow \\
    H_U = U\sum_{i} n_{i,\uparrow}n_{i,\downarrow}\\
    H_\sigma = -t \sum_{\langle ij\rangle} c^\dagger_{i,\sigma}c_{j,\sigma} - \mu_p\sum_i c^\dagger_{i,\sigma}c_{i,\sigma},
\end{gathered}
\end{equation}
where $U$ is the on-site Coulomb repulsion energy, taken to be the largest energy scale, $t$ is the hopping energy, $n_{i,\sigma} = c^\dagger_{i,\sigma}c_{i,\sigma}$ is the number of spin-$\sigma$ electrons on site $i$, and $\mu_p$ is a chemical potential, dependent on the doping $p$, that controls the total electron density. 
We consider only hole-doped systems with no net magnetization, so the number of electrons $N_e$ is less than the number of lattice sites $N$ and the numbers of spin-up and spin-down electrons are equal, $N_\uparrow = N_\downarrow = N_e/2$.
The number of doped holes is $N_h = N-N_e$, so the hole doping is $p = N_h/N$, and similarly the electron density is $n_e = N_e/N$.

\begin{figure*}
    \centering
    \includegraphics[width = 0.8\textwidth]{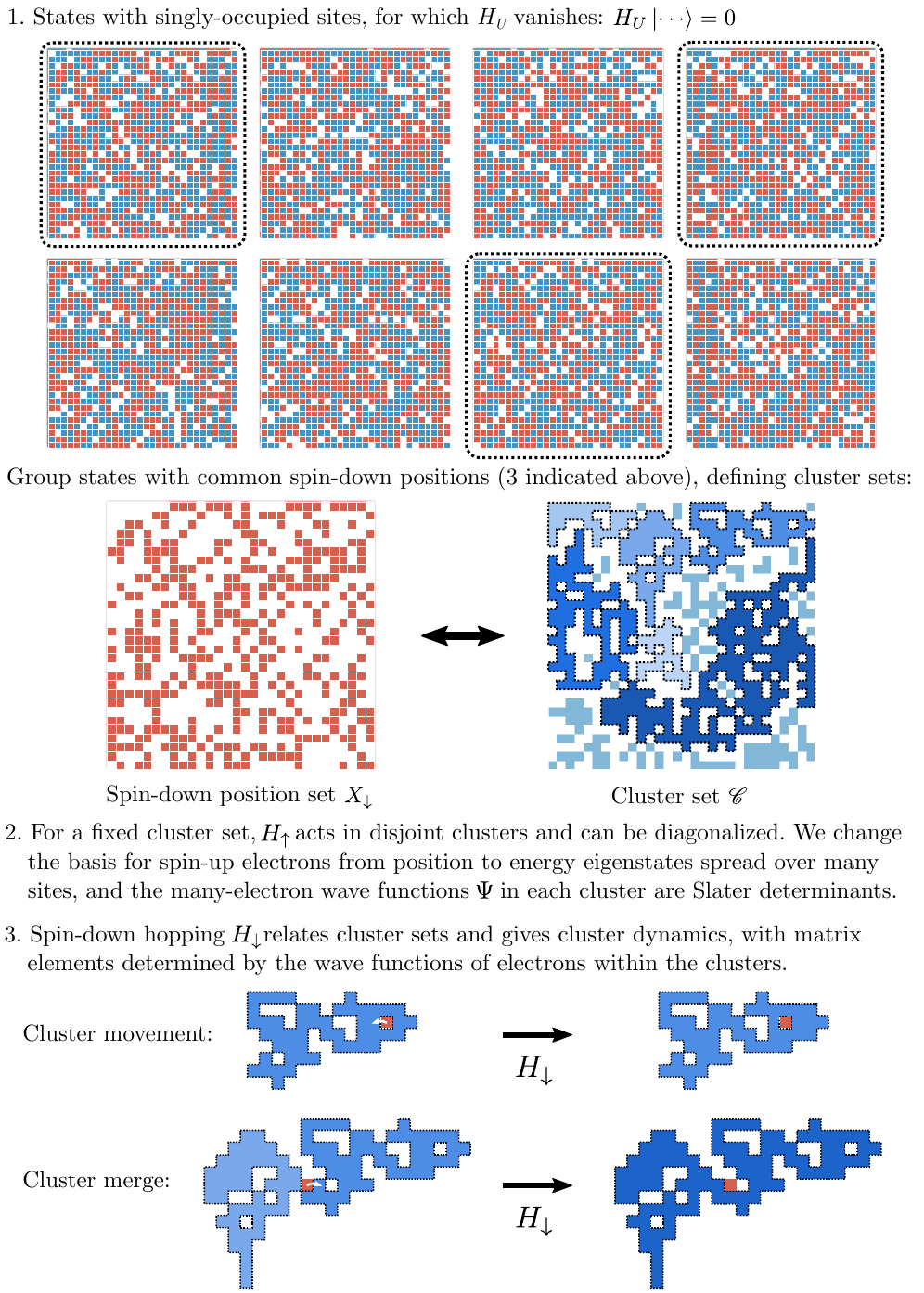}
    \caption{Schematic representation of process of rewriting the Hubbard model on the square lattice by implementing the percolation transition in the real-space many-electron basis states.
    The eight states shown at the top have only singly occupied sites, so that $H_U=0$, with spin-up (blue sites) and spin-down (red sites) electrons placed randomly. 
    The three states indicated with a dotted box have identical spin-down configuration $X_\downarrow$, shown below without spin-ups, which then uniquely defines a set of clusters of sites $\mathscr{C}$ where the spin-up electrons may reside.
    The six clusters larger than 20 sites in this example configuration are indicated.
    Spin-up electrons in the clusters are isolated, and $H_\uparrow$ can be diagonalized---the position basis transforms to the energy eigenbasis.
    Finally, we demonstrate how the hopping of a single spin-down electron via $H_\downarrow$ connects different cluster sets and produces dynamics in the clusters, such as simply shifting the center of mass of the sites in the cluster or merging clusters together.}
    \label{fig:schematic}
\end{figure*}

To build our model we consider the three terms of \cref{eq:HubbardH} in turn, and the full process is outlined schematically in \cref{fig:schematic}.
First, $H_U$ restricts the set of real-space basis states that we are free to use---large on-site repulsion $U\gg t,T$, essentially forbids doubly-occupied sites, so for a hole-doped system each site contains either a single electron or is empty, $n_e + p = 1$~\cite{Gutzwiller1965}. 
We can separate this Hilbert space into groups of states based on the positions of spin-down electrons; all states in a given group have spin-down electrons in the same positions, with only the positions of spin-up electrons differing between them.
Thus, states in each group are related to each other by spin-up electron hopping within fixed, disjoints clusters of available sites. 
Indeed, the set of these clusters of sites is in exact one-to-one correspondence with the positions of the spin-down electrons forming their boundaries, and either acts as a good label for states.

Next, focusing in on the level of a single cluster, we transform from the position basis for spin-up electrons to the eigenbasis of the spin-up hopping Hamiltonian $H_\uparrow$ within each cluster. 
Instead of definite position, spin-up electrons occupy states of definite energy with weight spread over many sites in a cluster, and many-electron states are given by Slater determinants of the corresponding wave functions.
We can write composite operators to create many-electron states in any cluster, and all basis states for the full system can be expressed in terms of these composite ``cluster operators.'' 

Finally, the remaining spin-down hopping term $H_\downarrow$ relates different spin-down electron configurations, equivalent to taking the system between different cluster sets. 
Since spin-down electrons form the boundaries of clusters, $H_\downarrow$ thus causes clusters to change in shape, merge, and divide. 
The states of spin-up electrons within the clusters are also changed, so matrix elements for spin-down electron hopping in this cluster language depend on the many-electron wave functions within the clusters. 
Taking all of these effects into account, we exactly rewrite the Hubbard Hamiltonian (for large $U$, projected into the single-occupancy low-energy sector) in terms of the composite cluster operators, which is then amenable to approximation and subsequent analytic exploration. 
\vspace{-12pt}
\subsection{\texorpdfstring{$H_U$}{Hubbard U} term --- Identify percolation transition} \label{sec:HU}
\vspace{-12pt}
The set of real-space basis states to describe the low-energy sector of the Hubbard model for large $U$ consists of all the ways of placing electrons into the system while avoiding double occupancy---a Gutzwiller projection~\cite{Gutzwiller1965}.
Let $X_\sigma$ be a set of $N_e/2$ sites hosting electrons of spin $\sigma$.
Then $X_\uparrow$ and $X_\downarrow$ have no intersection, and each is a proper subset of the other's complement, i.e. $X_\uparrow \subset\overline{X_\downarrow}$ and vice versa.
Any state in the basis is then specified by $X_\uparrow$ and $X_\downarrow$, and can be written as
\begin{equation} \label{eq:PsiXX}
    \ket{X_\uparrow,X_\downarrow} = \prod_{j\in X_\uparrow} c^\dagger_{j,\uparrow}\prod_{i\in X_\downarrow} c^\dagger_{i,\downarrow}\ket{0},
\end{equation}
where $\ket{0}$ is the empty-lattice vacuum state (cf. Eq. (4) in Ref.~\cite{Gutzwiller1965}). 

If we choose a state $\ket{X_\uparrow,X_\downarrow}$ from the low-energy Hilbert space at random in the thermodynamic limit then with probability $1$ the positions of the electrons in the state are uncorrelated up to the single-occupancy restriction.
Specifically, the probability that any site $\mathbf{x}$ is occupied by, say, a spin-down electron is given by spin-down electron density, $P(\mathbf{x}\in X_\downarrow) = n_e/2 = (1-p)/2$. 
Consequently, the probability that $\mathbf{x}$ \emph{does not} contain a spin-down electron is $P(\mathbf{x} \in \overline{X_\downarrow}) = 1-P(x\in X_\downarrow) = (1+p)/2$. 
The sites in $\overline{X_\downarrow}$ form clusters in the sense of a classical percolation problem as discussed in \cref{sec:percolation}, and the properties of these clusters are controlled by the ``occupation'' probability $P(\mathbf{x} \in \overline{X_\downarrow}) = (1+p)/2$, i.e. the probability that a site is not occupied by a spin-down electron. 
For small $P$ there are only finite clusters of ``not-down'' sites, and for large $P$ there is an infinite cluster of these sites. 
The critical point between these two regimes occurs at $P(\mathbf{x}\in\overline{X_\downarrow}) = P_c$, and we obtain a corresponding critical hole doping $p_c = 2P_c - 1$, giving
\begin{align}
        &p_c^{(\mathrm{sqr})} \approx 0.1854, \quad \text{square lattice} \\
        &p_c^{(\mathrm{hex})} \approx 0.3940, \quad \text{hexagonal lattice} \\
        &p_c^{(\mathrm{kag})} \approx 0.3054, \quad \text{kagome lattice},
\end{align}
using the values in \cref{tab:Pcvals}.
Though we have chosen spin-down electrons to bound the clusters here, breaking the symmetry between the model's treatment of the different spins, this choice is arbitrary and exchanging spin-up and spin-down everywhere yields the same results.
This argument depends on the hole doping only and is agnostic to the temperature of the system, so these values of $p_c$ are independent of temperature. 

For the triangular lattice with $P_c^{(\mathrm{tri})} = 1/2$ we obtain a critical doping $p_c^{(\mathrm{tri})} = 0$, i.e. half-filling, and while we could consider this lattice in principle, we would find no dynamics. 
As we will see, for a nontrivial rewriting of the low-energy sector of \cref{eq:HubbardH} the method developed here requires electrons to be able to hop without doubly-occupying sites, which is only possible with a finite density of doped holes. 
Thus we do not erroneously find metallic properties for systems known to be Mott insulators; for the triangular lattice the rewriting of the Hamiltonian would be entirely trivial. 

We see that the value of $p_c^{(\mathrm{sqr})}$ is rather close the critical hole doping $p_c\sim 0.19$ identified in experiments in a number of different cuprates~\cite{Loram2001,Tallon2001,Tallon2020}.
It is important to acknowledge, however, that experiments in some materials find a rather different value, e.g. perhaps as high as $p_c \sim 0.23$ in Nd-LSCO~\cite{Matt2015,Collingnon2017,Proust2019}. 
The value of the percolation critical occupation probability $P_c$ and therefore the corresponding hole doping $p_c$ is not universal, however, so perturbations to a system will change it in general. 
Perhaps most relevant in this respect is the spin-spin exchange interaction with energy scale $J\sim t^2/U$ produced for large but finite $U$ by virtual double occupation of sites, which could in principle produce long-range spin correlations for $T\lesssim J$~\cite{Dagotto1994}.
We continue with the classical percolation model with uncorrelated electron positions by arguing this term is only a very small perturbation for $J\ll t,T \ll U$, but a more general random cluster model~\cite{Fortuin1972, Grimmett2003}---which has percolation, Ising, and Potts models as special cases---may be more warranted and give more quantitatively accurate predictions for $T\sim J$, including a shift in nonuniversal properties such as the values of $p_c$. 
Even with this caveat, however, we will show that the qualitative behavior of the resistivity and optical conductivity of our system ultimately do not depend upon precise details of the percolation transition, e.g. the value of critical exponents, and only rely on the presence of a critical distribution of random clusters. 
Additionally, any disorder in the system, be it random spatial variation of the hopping $t$ or a random on-site potential, would also disfavor spatial correlations. 
Disorder is unavoidable in real systems, so the most relevant states to consider in our further analysis are precisely those without any order, manifesting the percolative behavior with $T$-independent transition that we have identified.
\vspace{-12pt}
\subsection{\texorpdfstring{$H_\uparrow$}{Spin-up hopping} term --- Cluster operators} \label{sec:Hup}
\vspace{-12pt}
To continue, we introduce some notation.
For any state written as in \cref{eq:PsiXX} with a given $X_\downarrow$, we define $\mathscr{C}$ to be the set of all clusters bounded by the sites in $X_\downarrow$---it is just $\overline{X_\downarrow}$ appropriately subdivided. 
(See the example in \cref{fig:schematic}.)
Since the sets $X_\downarrow$ and $\mathscr{C}$ are in one-to-one correspondence, we will replace $X_\downarrow\to\mathscr{C}$ in our labeling of states.  
We consider only $p\leq p_c$, so in the thermodynamic limit every $\mathscr{C}$ contains only finite clusters, which we denote as $\zeta\in\mathscr{C}$.
Spin-up electrons cannot hop between clusters, so for a fixed $\mathscr{C}$ all states $\ket{X_\uparrow, \mathscr{C}}$ are block-diagonal in $\zeta$ and we have
\begin{equation} \label{eq:stateclustersum}
    \ket{X_\uparrow,\mathscr{C}} = \bigoplus_{\zeta\in\mathscr{C}}\prod_{i\in X_{\uparrow,\zeta}}c^\dagger_{i,\uparrow}\ket{\varnothing,\zeta},
\end{equation}
where $X_{\uparrow,\zeta}$ are the positions of the spin-up electrons in cluster $\zeta$ (the union of all $X_{\uparrow,\zeta}$ is $X_\uparrow$) and $\ket{\varnothing,\zeta}$ is the state with all sites in the cluster unoccupied. 

It is useful as we continue to have a way to uniquely specify any cluster $\zeta$. 
One way is to list the positions of all of the sites comprising it, but this is cumbersome and will not be a useful strategy. 
Instead we label each cluster first with the number of sites comprising it $s$ and the center of mass of those sites $\mathbf{r} = \sum_{i=1}^s \mathbf{r}_i/s$, where $\mathbf{r}_i$ are the positions of the $s$ sites in the cluster, so that any $\mathbf{r}$ can be written as some integer multiples of $\mathbf{a}_j/s$, where $\mathbf{a}_j$ are the vectors needed to specify the position of each lattice site---two lattice vectors to locate unit cells, and an additional vector for each lattice site in each unit cell beyond the first.
For a given $s$, all allowed $\mathbf{r}$s in the square lattice also form a square lattice, and in the hexagonal and kagome lattices the allowed $\mathbf{r}$s form a regular triangular lattice.
In all cases, the closest allowed positions are separated from each other by a distance $a/s$. 
Finally, let the index $\lambda$ label all possible distinct shapes a cluster may have for a given $s$. 
We will denote the total number of these shapes as $N_{\lambda}(s)$, so that $\lambda = 1, 2,\dots,N_{\lambda}(s)$ for a given $s$. 
For even relatively small $s$ this number is large, e.g. in the square lattice $N_\lambda(10)=36446$ and $N_\lambda(24)\sim 10^{13}$~\cite{Redelmeier1981}.
The triples $\{s,\mathbf{r},\lambda\}$ then uniquely specify each cluster---they are sufficient information to determine the positions of all $s$ sites---and two clusters $\zeta = \{s,\mathbf{r},\lambda\}$ and $\zeta' = \{s',\mathbf{r}',\lambda'\}$ are identical if and only if $s=s'$, $\mathbf{r}=\mathbf{r}'$, and $\lambda=\lambda'$.
Sums over clusters can then be understood as summation over $s$, and for each $s$ appropriate sums over $\mathbf{r}$ and $\lambda$. 

In the basis of states written as \cref{eq:stateclustersum}, the spin-up hopping term of \cref{eq:HubbardH} becomes
\begin{equation}
    H_\uparrow = -t\sum_{\langle ij\rangle}c^\dagger_{i,\uparrow}c_{j,\uparrow} -\mu_p \sum_{i}c^\dagger_{i,\uparrow}c_{i,\uparrow} =  \sum_{\zeta\in\mathfrak{C}} H_\zeta
\end{equation}
where $\mathfrak{C}$ is the set of all finite clusters appearing in \emph{any} $\mathscr{C}$, and $H_\zeta$ is the Hamiltonian for spin-up electrons hopping inside cluster $\zeta$.
We can diagonalize each $H_\zeta$ individually,
\begin{equation}
    H_\zeta = -t\sum_{\langle ij\rangle\in\zeta} c_{i,\uparrow}^\dagger c_{j,\uparrow} -\mu_p\sum_{i\in\zeta} c_{i,\uparrow}^\dagger c_{i,\uparrow} = \sum_{i=1}^s \epsilon_{\zeta,i} \psi_{\zeta,i}^\dagger \psi_{\zeta,i},
\end{equation}
where $\psi_{\zeta,i}$ ($\psi_{\zeta,i}^\dagger$) is the operator that destroys (creates) a spin-up electron in cluster $\zeta$ with energy $\epsilon_{\zeta,i}$, and $s$ is the number of sites in the cluster.
(For simplicity we have dropped the spin label in this basis.) 
These $\psi$'s are related to the $c_\uparrow$'s as
\begin{equation}
    c_{\mathbf{x},\uparrow} = \sum_{i=1}^s \varphi_{\zeta,i}(\mathbf{x}) \psi_{\zeta,i},
\end{equation}
where $\mathbf{x}$ is the position of any of the $s$ sites in $\zeta$ and $\varphi_{\zeta,i}(\mathbf{x})$ are the wave functions for the single-particle fermionic states in the cluster. 
We use the convention that $\epsilon_{\zeta,i} \leq \epsilon_{\zeta,j}$ if $i<j$. 
The energies of these single-particle states and their corresponding wave functions depend on the full geometry of the cluster, but because this is a subset of the full two-dimensional lattice we know that $\abs{\epsilon_{\zeta,i}}$ is bounded by a scale of order $t$ for all clusters. 

We can now write multielectron states in each cluster with well-defined total energy, and then use these to write a basis set for the whole system, in which $H_\uparrow$ is diagonal for any fixed $\mathscr{C}$. 
We denote the number of electrons within cluster $\zeta$ as $\nu_\zeta$ $(\leq s)$, and the indices of the occupied states are collected in the set $\mathcal{I}_\zeta = \{i_1, i_2,\cdots,i_{\nu_\zeta}\}$, ordered by their corresponding energies. 
Every possible $\mathcal{I}_\zeta$ is an element of the power set of the $s$ total indices $\mathbb{P}_s$. 
multifermion energy eigenstates in the cluster are
\begin{multline}
    \ket{\mathcal{I}_\zeta,\zeta} = \psi_{\zeta,i_{\nu_\zeta}}^\dagger\cdots\,\psi^\dagger_{\zeta,i_1} \ket{\varnothing,\zeta} \\
    = \sum_{\{\mathbf{x}_i\}\in\zeta} \!\!\Phi_{\zeta,\mathcal{I}_\zeta}(\mathbf{x}_1,\dots,\mathbf{x}_{\nu_\zeta}) \,c^\dagger_{\mathbf{x}_{\nu_\zeta},\uparrow} \cdots \,c^\dagger_{\mathbf{x}_1,\uparrow}\ket{\varnothing,\zeta},
\end{multline}
A multifermion wave function is given by the Slater determinant
\begin{equation} \label{eq:clusterWF}
    \Phi_{\zeta,\mathcal{I}_\zeta}(\mathbf{x}_1\dots\mathbf{x}_{\nu_\zeta}) = \frac{1}{\sqrt{\nu_\zeta!}} \abs{\begin{array}{ccc}
        \varphi_{\zeta,i_1}(\mathbf{x}_1)  & \cdots & \varphi_{\zeta,i_{\nu_\zeta}}(\mathbf{x}_1) \\
        \vdots & \ddots & \vdots \\
        \varphi_{\zeta,i_1}(\mathbf{x}_{\nu_\zeta}) & \cdots & \varphi_{\zeta,i_{\nu_\zeta}}(\mathbf{x}_{\nu_\zeta}) 
    \end{array}},
\end{equation} 
and the total energy of the state is
\begin{equation}
    E_{\zeta,\mathcal{I}_\zeta} = \sum_{i\in\mathcal{I}_\zeta} \epsilon_{\zeta,i}.
\end{equation}

We write a multifermion composite operator to create these states,
\begin{equation}
    C_{\zeta,\mathcal{I}_\zeta}^\dagger = \mathcal{P}_{\zeta,\mathcal{I}_\zeta} \psi^\dagger_{\zeta,i_{\nu_\zeta}}\cdots\,\psi^\dagger_{\zeta,i_1},
\end{equation}
using the projector
\begin{equation}
    \mathcal{P}_{\zeta,\mathcal{I}_\zeta} = \prod_{i\in\mathcal{I}_\zeta} \psi^\dagger_{\zeta,i} \psi_{\zeta,i} \prod_{j\not\in\mathcal{I}_\zeta}  \psi_{\zeta,j} \psi^\dagger_{\zeta,j} \prod_{\mathbf{x}\in \partial\zeta} c^\dagger_{\mathbf{x},\downarrow} c_{\mathbf{x},\downarrow},
\end{equation}
where $\partial\zeta$ are the sites bounding cluster $\zeta$ occupied by spin-down electrons, which ensures that $C_\zeta$ only acts in cluster $\zeta$ and that the electrons are only put into the states indexed in $\mathcal{I}_\zeta$.
The corresponding annihilation operator $C_{\zeta,\mathcal{I}_\zeta}$ can be similarly defined, and together these operators act on cluster states as
\begin{gather}
    C^\dagger_{\zeta,\mathcal{I}_\zeta} \ket{\varnothing,\zeta'} = \delta_{\zeta,\zeta'} \ket{\mathcal{I}_\zeta,\zeta} \\
    C_{\zeta,\mathcal{I}_\zeta} \ket{\mathcal{I}'_{\zeta'},\zeta'} = \delta_{\mathcal{I}_\zeta,\mathcal{I}'_\zeta} \delta_{\zeta,\zeta'} \ket{\varnothing,\zeta},
\end{gather}
where the delta-function $\delta_{\zeta,\zeta'}$ is $1$ if $\zeta$ and $\zeta'$ are the same cluster and $0$ otherwise, and $\delta_{\mathcal{I}_\zeta,\mathcal{I}'_{\zeta'}}$ is $1$ if the two sets are identical and $0$ otherwise. 
The elements of the many-electron basis with well-defined total energy (as opposed to well-defined position as in \cref{eq:stateclustersum}) can be then be written
\begin{equation}
    \ket{\left\{\mathcal{I}\right\},\mathscr{C}} = \bigoplus_{\zeta\in\mathscr{C}} C^\dagger_{\zeta,\mathcal{I}_\zeta} \ket{\varnothing,\zeta},
\end{equation}
where the spin-up electron position set $X_\uparrow$ is replaced with $\left\{\mathcal{I}\right\}$, the set of occupation sets for the clusters. 
The Hamiltonian in each cluster $H_\zeta$ is then
\begin{equation} 
    H_\zeta = \sum_{\mathcal{I}_\zeta\in\mathbb{P}_s} E_{\zeta,\mathcal{I}_\zeta} C^\dagger_{\zeta,\mathcal{I}_\zeta} C_{\zeta,\mathcal{I}_\zeta},
\end{equation} 
so that the spin-up hopping term of the Hubbard Hamiltonian is 
\begin{equation} \label{eq:uphopping}
    H_{\uparrow} = \sum_{\zeta\in\mathfrak{C}} \sum_{\mathcal{I}_\zeta\in\mathbb{P}_s} E_{\zeta,\mathcal{I}_\zeta} C^\dagger_{\zeta,\mathcal{I}_\zeta} C_{\zeta,\mathcal{I}_\zeta}.
\end{equation}

Since the total spin-up electron density is fixed to $n_e/2$, the average filling of each cluster is
\begin{equation} \label{eq:filling}
    \expval{\nu_\zeta} = \frac{\sum_{\zeta\in\mathscr{C}}\nu_\zeta}{\sum_{\zeta\in\mathscr{C}}s} = \frac{\frac{N_e}{2}}{N-\frac{N_e}{2}} = \frac{\frac{n_e}{2}}{1-\frac{n_e}{2}} = \frac{1-p}{1+p} \equiv \rho_p.
\end{equation}

Notice that the parity of fermionic operators comprising $C_{\zeta,\mathcal{I}_\zeta}$ or $C^\dagger_{\zeta,\mathcal{I}_\zeta}$ is determined by the number of electrons within the cluster $\nu_\zeta$. 
If $\nu_\zeta$ is even, then these operators are bosonic, and if $\nu_\zeta$ is odd they are fermionic, and we can define
\begin{equation}
    C_{\zeta,\mathcal{I}_\zeta} \equiv \begin{cases}
    b_{\zeta,\mathcal{I}_\zeta}, & \nu_\zeta \text{ even},\\
    f_{\zeta,\mathcal{I}_\zeta}, & \nu_\zeta \text{ odd},
    \end{cases}
\end{equation}
to make this difference explicit.
\vspace{-12pt}
\subsection{\texorpdfstring{$H_\downarrow$}{Spin-down hopping} term --- Cluster dynamics} \label{sec:Hdown}
\vspace{-12pt}
Spin-down electrons define cluster boundaries, so when projected into the cluster basis the spin-down hopping term of \cref{eq:HubbardH} describes dynamics of the clusters themselves.  
A single spin-down hop may simply shift the center of mass (and change the shape) of a single cluster, it may cause multiple clusters to merge, or cause one cluster to divide. 
More concretely, the spin-down hopping term provides matrix elements in the many-body Hamiltonian between states labeled by different cluster sets $\mathscr{C}$ and $\mathscr{C}'$ that differ only in the position of a single spin-down electron.
Since only a single spin-down has moved between $\mathscr{C}$ and $\mathscr{C}'$, they are the same up to only a small number of clusters; the number of clusters that can share a site on all of their boundaries is given by the coordination number of the lattice, so all of the clusters in $\mathscr{C}$ and $\mathscr{C}'$ are identical with only a finite number of exceptions---four in the square and kagome lattices, three in the hexagonal lattice. 

For a spin-down electron at position $\mathbf{x}$, let $N_c$ be the number of clusters in $\mathscr{C}$ with this electron forming part of their boundary, which we will label $\zeta_1,\dots,\zeta_{N_c}$.
If this electron hops to position $\mathbf{x}+\mathbf{e}$, where $\mathbf{e}$ is a nearest-neighbor vector, then it is now on the boundary of $N'_c$ new clusters in set $\mathscr{C}'$, labeled $\zeta'_1,\dots,\zeta'_{N'_c}$. 
All other clusters are unaffected, so the spin-down hopping term of the Hubbard Hamiltonian projected into the cluster basis is simply $-t$ times the inner product of the initial and final cluster states, 
\begin{equation} \label{eq:gdef}
     g^{(N_c\to N'_c)}_{\alpha'_1\dots\alpha'_{N'_c}; \alpha_1\dots\alpha_{N_c}} \equiv \bra{\left\{\mathcal{I}'\right\}\!,\mathscr{C}'} \!\Big(\sum_{\langle ij\rangle}c^\dagger_{i,\downarrow} c_{j,\downarrow} \Big)\! \ket{\left\{\mathcal{I}\right\}\!,\mathscr{C}},
\end{equation}
where we have also introduced the shorthand notation $\alpha_i = \zeta_i,\mathcal{I}_{\zeta_i}$ with associated sums
\begin{equation}
    \sum_{\alpha_i} \cdots = \sum_{\zeta_i\in \mathfrak{C}} \sum_{\mathcal{I}_{\zeta_i}\in\mathbb{P}_{s_i}} \cdots.
\end{equation}
This inner product can be expressed in terms of the multielectron wave functions  \cref{eq:clusterWF} for the involved clusters.
Since the wave functions strongly depend on cluster shapes, these matrix elements do as well---all other indices being fixed, changing even just one of the shape indices $\lambda_1,\dots,\lambda_{N_c}$ or $\lambda'_1,\dots,\lambda'_{N'_c}$ can significantly affect the value of the matrix element.
The initial and final sets of clusters are composed of the same number of sites $s = s_1 + \cdots + s_{N_c}$, and the positions of these sites are the same with a single exception, so the total center of mass of the clusters shifts a distance $a/s$ in the direction opposite to the spin-down hop. 
(See the examples in \cref{fig:schematic}.)
We give an explicit calculation of this matrix element in \cref{app:gcalculation} incorporating all constraints.

The spin-down electron chemical potential term is much less complicated. 
No electrons are moved by this term, so in cluster language it contributes to the terms diagonal in cluster configuration space like the spin-up hopping term. 
\vspace{-12pt}
\subsection{\texorpdfstring{$H$}{H} in terms of clusters} \label{sec:rewriteH}
\begin{figure}[!ht]
    \centering
    \includegraphics[width=\columnwidth]{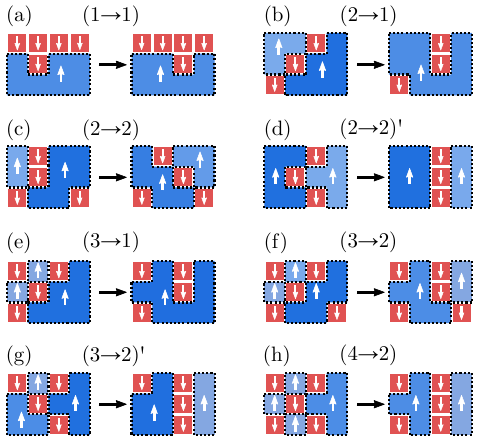}
    \caption{Representatives of eight distinct cluster interactions produced by the single hop of one spin-down electron (red squares with $\downarrow$) in the square lattice, excluding processes that include empty single-site clusters.
    The blue regions marked with $\uparrow$ represent clusters containing the spin-up electron wave functions $\Phi$ as in \cref{eq:clusterWF}.
    Processes labeled (b), (e), (f), (g), and (h) reduce the number of clusters in the system, and each have a corresponding time reversed process that increases the number of clusters. 
    Interactions with a primed $(N_c\to N'_c)$ label differ from their unprimed counterparts by containing at least one cluster that does not exchange spin-up electrons with other clusters---its size changes by 1, but the number of occupied states within it cannot. 
    Examples of panels (a) and (b) for complete clusters are also shown in \cref{fig:schematic}.
    Because the hexagonal lattice has a lower coordination number, there are fewer possible types of interactions than listed here.}
    \label{fig:clusterdynamics}
\end{figure}

\vspace{-12pt}
We can now restate the Hubbard Hamiltonian \cref{eq:HubbardH} for large $U$ in terms of clusters,
\begin{align} \label{eq:clusterH}
    H &= \sum_\alpha E_\alpha C^\dagger_\alpha C_\alpha - \mu_p \sum_{\{\alpha_i\}} C^\dagger_{\alpha_1} \cdots C^\dagger_{\alpha_{N_c}} C_{\alpha_{N_c}} \cdots C_{\alpha_1} \nonumber \\
    &-t \!\!\! \sum_{\{(N_c\to N'_c)\}} \sum_{\{\alpha'_i\}} \sum_{\{\alpha_i\}} g^{(N_c\to N'_c)}_{\alpha'_1\dots\alpha'_{N'_c};\alpha_1\dots\alpha_{N_c}}  \nonumber \\
    & \hspace{3cm}\times C^\dagger_{\alpha'_1} \cdots C^\dagger_{\alpha'_{N'_c}} C_{\alpha_1} \cdots C_{\alpha_{N_c}}.
\end{align}
The first term is the entire spin-up hopping $H_\uparrow$, the second term accounts for the chemical potential part of $H_\downarrow$---the sets of $N_c$ clusters being summed over are those which share a spin-down electron on their boundary---and the final term gives the effect of spin-down electron hopping.
For this to be a complete rewriting we need to identify the different allowed cluster transformations $(N_c\to N'_c)$ generated by the spin-down hopping term. 
As noted above, we are guaranteed that there are only a finite number of these since $N_c$ and $N'_c$ are bounded from above. 
In \cref{fig:clusterdynamics} we demonstrate the 8 distinct types of processes in the square lattice, ignoring cases where one of the involved clusters has $s=1$ and $\nu=0$. 
(We will keep only large clusters in our later approximations). 
None involve more than six total clusters--four becoming two or vice versa--so no term in \cref{eq:clusterH} for the square lattice can involve more than six cluster operators.
For hexagonal and kagome lattices there are similar restrictions on the total number of possible operators; these have a coordination number of 3, so a term in the Hamiltonian can contain at most five clusters---three initial clusters becoming two final cluster, or vice versa. 
Even though \cref{eq:clusterH} comes entirely from electron kinetic energy terms, we will refer to the multicluster terms as interactions since they have the form of interactions between clusters. 

Notice that despite the Hubbard Hamiltonian being completely local, the interaction matrix elements depend on the positions of electrons that are far apart; the cluster wave functions depend on the full shape of the involved clusters.
This reflects the nonlocality of the strong correlations in the system generated by the Hubbard $U$.

\section{Approximations} \label{sec:approximations}
\vspace{-12pt}
Since \cref{eq:clusterH} is equivalent to the original infinite-$U$ Hubbard Hamiltonian it is at least as difficult to exactly calculate its properties, and to continue we must make approximations. 
The advantage of the rewriting is in the ability to make these approximations.
The first is to keep only the Gaussian $(1\to1)$ and and cubic $(2\to1)$ and $(1\to 2)$ interaction terms, i.e. the $C^\dagger C$, $C^\dagger CC$, and $C^\dagger C^\dagger C$ terms. 
These two interactions are enough to generate all qualitatively distinct cluster dynamics---moving, changing shape, merging, and dividing. 
Because they require far less strict local configurations of electrons, these two interactions are far more likely to occur than any of the others, a fact supported by simple numerical simulations counting the prevalence of these local arrangements in random spin-down electron configurations for $p\sim p_c$ in the square lattice. 
Additionally, local configurations are not sufficient to determine the actually number of clusters affected by a particular spin-down hop---clusters that seem distinct in the local vicinity of the relevant boundary electron may join up further away. 
Therefore, interactions involving more clusters are even less likely than found with a naive counting of local configurations. 
\vspace{-24pt}
\subsection{Large clusters, IR limit}
\vspace{-12pt}
As noted in \cref{sec:percolation} the critical properties of the percolation transition manifest in the scaling behavior of large clusters, those with sizes above some $s_0\gg1$, corresponding to a linear size larger than $R_0\gg a$.
Therefore, considering only the physics of the system on large length scales--an infrared or continuum approximation--is equivalent to focusing on large clusters, and so preserves the critical properties of large clusters while also allowing for further simplification. 

Consider the effects of %a restriction to large cluster 
this approximation on the quantities introduced to label clusters.
First, sums over all discrete sizes $s$ can be replaced with integrals.
This approximation works very well for large clusters, as is most easily seen by changing from integer-valued $s$ to real-valued $R \propto s^{1/d_f}$.
Second, the number of possible cluster shapes $N_\lambda(s)$ is incredibly large for large $s$, so anticipating that we will drop terms suppressed by inverse powers of this number, we replace $N_\lambda(s)$ with a uniform large $N_\lambda$ for all $s$ to simplify intervening calculation. 
Finally, the spacing between the possible center-of-mass locations is $a/s$, so for large clusters the allowed values of $\mathbf{r}$ may be well-approximated as a continuum. 
It is simplest to implement this continuum approximation after transforming to reciprocal space---long wavelengths $R \gg a/s$ correspond to small momenta $ka/s \ll 1$, allowing straightforward expansions. 

For large $s$ we can also make approximations to the states of spin-up electrons within the clusters. 
Numerical evaluation of the single-particle spectra for large random clusters in the square lattice shows that they are well approximated by averages over shape; if $\bar{\epsilon}_{s,i}$ is the average of the energies over $\lambda$ for a given $s$, then $\abs{\epsilon_{\mathbf{r},s,\lambda,i}-\bar{\epsilon}_{s,i}} \ll t$.
Furthermore, the average energies for different $s$ all appear to be discrete samplings of the same continuous and gapless function $\bar{\epsilon}(x)$ with $x\in[0,1]$ indexing a continuum of states. 
(See \cref{sec:clusterspectra} for this numerical analysis.)
Altogether we put
\begin{equation}
    \epsilon_{\mathbf{r},s,\lambda,i} \approx \bar{\epsilon}(i/(s+1)) \widesim{s\gg1} \bar{\epsilon}(x).
\end{equation}
All dependence of these energies on cluster size, shape, and position are dropped. 

In the IR limit the sets $\mathcal{I}$ labeling discrete occupied states can be approximated as occupation functions $n(x)$ defined on the continuum of internal states. 
When integrated over all $x$, the occupation function gives the fraction of occupied states, $\rho = \int \dd x\, n(x) \approx \nu/s$.
To simplify further we assume that the physically relevant case for describing the low-energy properties of the system is that of thermal occupation, so that $n(x) = n_F(\bar{\epsilon}(x)-\mu)$, where $n_F(\epsilon) = \left(1+e^{\epsilon/T}\right)^{-1}$ is the Fermi function and $\mu = \bar{\epsilon}(\rho)$ so that $\rho$ completely characterizes the occupation of the internal states. 
Since the addition or removal of a single electron does not change the filling fraction in the continuum approximation, we have both a bosonic and fermionic cluster for any $\rho$. 
Finally, we will assume that all clusters have the same fraction of their states occupied, which, with the above approximations, means they are filled to the same energy, i.e. we take $\rho = \rho_p = (1-p)/(1+p)$ in all clusters, the average filling found in \cref{eq:filling}, so that $\mu = \mu_p$ in all clusters as well. 
At the critical doping this filling is $\rho_{p_c^{(\mathrm{sqr})}} \approx 0.687$ for the square lattice, $\rho_{p_c^{(\mathrm{hex})}} \approx 0.435$ in the hexagonal lattice, and $\rho_{p_c^{(\mathrm{kag})}} \approx 0.532$ for the kagome lattice.
The total internal energy of a cluster is then
\begin{align}
    E_{\mathbf{r},s,\lambda,\mathcal{I}} &= \sum_{i\in\mathcal{I}}\epsilon_{\mathbf{r},s,\lambda,i}
    \widesim{s\gg 1} s \int_0^1 \dd x\,n_F\left(\bar{\epsilon}(x)-\bar{\epsilon}(\rho_p)\right)\,\bar{\epsilon}(x) \nonumber \\
    &\equiv s\, E_p. 
\end{align}
\vspace{-28pt}
\subsection{Interaction Matrix Elements}
\vspace{-12pt}
Averaging over cluster shapes gives an approximation of the internal spectra of clusters, but averaging over shapes is a meaningless operation for the wave functions within clusters. 
The wave functions of internal states are intrinsically related to cluster geometry and there are no ``average wave functions'' that can approximate the interaction matrix elements $g^{(N_c\to N'_c)}_{\alpha'\dots;\alpha\dots}$.
Therefore, though $\lambda$ is an irrelevant index at the level of single-cluster properties, any treatment of the interactions must retain strong dependence on this degree of freedom.

First note that for large $s$ interactions are mostly independent of $s$. 
Because of the same geometric properties that lead to the fractal relationship between cluster radius $R$ and size $s$, wave functions are typically localized within some length $\ell$ set by the lattice scale~\cite{Wang1995, Konig2012, vanVeen2016, Kempkes2019}, and moving a single site in a cluster will affect only the wave functions with weight near that site. 
Since the interaction matrix elements are the inner product of the wave functions in clusters that differ by the location of a single site, only a small region of scale $\ell$ determines the size of any interaction amplitude. 
Since $R\gg \ell\sim a$ for the large clusters we consider, then the matrix elements do not strongly depend on cluster size.  

With the approximations we have discussed so far the $(1\to1)$ interaction matrix element becomes 
\begin{equation} \label{eq:11matelement}
    g^{(1\to1)}_{\alpha;\alpha'} \approx  g_{\lambda\lambda'}\, \delta_{s,s'}\sum_{\mathbf{e}}\delta(\mathbf{r}-\mathbf{r}'-\mathbf{e}/s),
\end{equation}
where $g_{\lambda\lambda'}$ is independent of cluster position and size and encodes all shape dependence. 
The $\delta$-function here gives the shift of the center of mass of the cluster---the final position $\mathbf{r}$ is shifted from the initial position $\mathbf{r}'$ by a vector of length $a/s$ opposite the direction of the electron hop.
In the square lattice electrons may hop along the $x$ or $y$ directions so that $\mathbf{e} \in \{\pm a\hat{\mathbf{x}},\pm a\hat{\mathbf{y}}\}$.
We orient the hexagonal and kagome lattices to have one link along $x$, so that the others are rotated from this by $\pm\pi/3$ and $\mathbf{e} \in \{\pm a\hat{\mathbf{x}}, \pm a(\hat{\mathbf{x}}/2 + \hat{\mathbf{y}} \sqrt{3}/2), \pm a(\hat{\mathbf{x}}/2 - \hat{\mathbf{y}} \sqrt{3}/2)\}$; though the lattice structure is different, the allowed directions for nearest-neighbor electron hopping, and therefore shifts in cluster center-of-mass, are the same in these two lattice.  

A cluster can only change into a small fraction of the $N_\lambda$ possible shapes by moving a single boundary electron, so for a fixed $\lambda$ only a relatively small number of $\lambda'$ give nonzero result. 
The square of the matrix elements is nonnegative, so their total is some very small positive number,
\begin{gather}
    \sum_{\lambda'} \abs{g_{\lambda\lambda'}}^2 \equiv g_{2,\lambda}^2,
\end{gather}
which will depend on the starting cluster shape $\lambda$. 
At this level we can now consider the effect of averaging over shape.
The nonzero values of $g_{\lambda\lambda'}$ are just as likely to be positive as negative, so its average over shapes vanishes. 
We use $g_2^2$ for the average value of $g_{2,\lambda}^2$ over the $N_\lambda$ initial cluster shapes $\lambda$, so averaging over shape indices gives
\begin{equation}
    \expval{g_{\lambda\lambda'}} = 0 \qquad \expval{\abs{g_{\lambda\lambda'}}^2} = \frac{g_2^2}{N_\lambda},
\end{equation}
where $\expval{\dots}$ denotes averaging over any unsummed shape indices. 
The small constant $g_2^2$ giving the average probability for a transition from one cluster shape to any other therefore characterizes the strength of this interaction term. 

The same considerations apply directly for the cubic $(2\to1)$ interaction, but with one complication---not all position dependence is removed by our approximations, and the matrix element still depends on the separation between the two initial clusters.
For very large separations, much larger than the sum of the two cluster radii, two clusters are very unlikely to share a boundary so the corresponding amplitude vanishes.
For very small separations they are likely to overlap, meaning they cannot both exist at once and likewise the corresponding amplitude vanishes. 
Only in between, in a wide window around the sum of the radii, are there separations with nonzero amplitude. 
Since the average cluster radius diverges near $p_c$ we will extend this window up to the size of the system, and in the IR limit let the interaction be independent of this separation as well; for the most important large clusters near $p_c$, the interaction appears maximally nonlocal in real space.
We therefore have 
\begin{equation} \label{eq:21matelement}
    g^{(2\to1)}_{\alpha';\alpha_1,\alpha_2} \approx g_{\lambda';\lambda_1\lambda_2} \delta_{s_1+s_2,s'} \sum_{\mathbf{e}} \delta\left(\frac{s_1}{s}\mathbf{r}_1+\frac{s_2}{s}\mathbf{r}_2-\mathbf{r}'-\frac{\mathbf{e}}{s}\right)
\end{equation}
with $g_{\lambda';\lambda_1\lambda_2}$ encoding all shape dependence of the matrix element. 
As in \cref{eq:11matelement}, the $\delta$-function gives the shift in the total center of mass position between the initial and final clusters.  
Similar arguments about the properties of this quantity and the nature of the average over shapes as given for $g_{\lambda\lambda'}$ above now give
\begin{equation}
    \expval{g_{\lambda';\lambda_1\lambda_2}} = 0 \qquad
    \expval{\abs{g_{\lambda';\lambda_1\lambda_2}}^2} = \frac{g_3^2}{N_\lambda^2},
\end{equation}
where $g_3^2$ is a positive constant characterizing the strength of this interaction. 
There are many fewer ways to satisfy all of the constraints of the $(2\to1)$ interaction than for the $(1\to1)$ interaction, so there are fewer nonzero values of $g_{\lambda';\lambda_1\lambda_2}$ than $g_{\lambda\lambda'}$, and $g_3^2 \ll g_2^2$. 

We see that our approximations leave us with interaction constants that depend on ``flavor'' indices which take a large number of values and are characterized by particular statistical properties after an averaging procedure. 
This looks very similar to the structure posited in large-$N$ models with random interactions such as the SYK and Yukawa-SYK models~\cite{Chowdhury2022}.
The quenched-disordered interactions in those models are self-averaging for many quantities, so considering a random fixed set of couplings gives the same result as averaging over realization of the disorder. 
Similarly, the specific fixed interactions that manifest in a randomly chosen state in our model are also random, and the average over cluster shapes can simply be thought of as a way to calculate quantities that are self-averaging. 
\vspace{-24pt}
\subsection{Approximate Hamiltonian}
\vspace{-12pt}
Implementing all of the above approximations into \cref{eq:clusterH}, writing cluster operators explicitly as either fermionic or bosonic, and then Fourier transforming real-space cluster positions to momentum space gives
\begin{widetext}
\begin{equation} \label{eq:approxH}
\begin{aligned}
    H &\approx \int_{0^+}^{s_\xi} \!\!\dd s \sum_{\mathbf{k}} \sum_{\lambda,\lambda'}^{N_\lambda} \left(s\,E_p\,\delta_{\lambda\lambda'}+t\,\phi(\mathbf{k})\,g_{\lambda\lambda'}\right) C^\dagger_\lambda(\mathbf{k},s) C_{\lambda'}(\mathbf{k},s) \\
    & \hspace{1cm}- \mu_p \sum_{\mathbf{k}} \sum_{N_c=1}^{4}M_{N_c}\prod_{i=1}^{N_c} \int_{0^+}^{s_\xi} \!\!\dd s_i \sum_{\mathbf{k}_i} \sum_{\lambda_i}^{N_\lambda} C^\dagger_{\lambda_i}(\mathbf{k}_i,s_i) C_{\lambda_i}(\mathbf{k}_i,s_i) \delta(\mathbf{k}-\mathbf{k}_1 \cdots - \mathbf{k}_{N_c}) \\
    & + \int_{0^+}^{s_\xi}\dd s \int_{0^+}^{s} \dd s' \sum_{\mathbf{k}} \sum_{\lambda,\lambda_1,\lambda_2}^{N_\lambda}t\,\phi(\mathbf{k}) \left[ g^{(f;fb)}_{\lambda;\lambda_1\lambda_2} \, f^\dagger_\lambda(\mathbf{k},s) f_{\lambda_1}\big(\mathbf{k}, s-s'\big) b_{\lambda_2}\big(\mathbf{k}, s'\big) \right. \\
    & \hspace{2cm}\left. + \frac{1}{2} g^{(b;ff)}_{\lambda;\lambda_1\lambda_2} \, b^\dagger_\lambda(\mathbf{k},s) f_{\lambda_1}\big(\mathbf{k}, s-s'\big) f_{\lambda_2}\big(\mathbf{k}, s'\big) + \frac{1}{2} g^{(b;bb)}_{\lambda;\lambda_1\lambda_2} \, b^\dagger_\lambda(\mathbf{k},s) b_{\lambda_1}\big(\mathbf{k}, s-s'\big) b_{\lambda_2}\big(\mathbf{k}, s'\big)\right] + \mathrm{h.c.},
\end{aligned}
\end{equation}
\end{widetext}
where the function $\phi(\mathbf{k})$ result from the Fourier transform and the $\delta$-functions in \cref{eq:11matelement,eq:21matelement} constraining the total cluster center-of-mass to shift along the allowed directions.
For the square lattice it has the form
\begin{equation}
    \phi(\mathbf{k}) = -2\left[\cos\left(k_x a\right) + \cos\left(k_y a\right)\right] \approx k^2a^2 -4
\end{equation}
and for the hexagonal and kagome lattices it is
\begin{multline}
    \phi(\mathbf{k}) = -2\left[\cos\left(k_x a\right) + 2\cos\left(\frac{k_x a}{2}\right)\cos\left(\frac{\sqrt{3}k_y a}{2}\right)\right] \\
    \approx \frac{3}{2}k^2a^2 - 6,
\end{multline}
with expansions in both cases valid for $\abs{ka}\ll 1$.
The constants $M_{N_c}$ are the average number of boundary spin-down electrons shared by $N_c$ clusters. 

The interactions here are local in momentum space, reflecting the approximations made to the interaction coefficients above.
The interaction between clusters is nonzero when the separation between their centers of mass lies within a window around their radii. 
As the typical cluster radius diverges near $p_c$, so to does the range over which this interaction extends, leading to an apparently entirely nonlocal interaction in real space, and therefore completely local form in momentum space.

Obtaining this form of the Hamiltonian involves a redefinition of the cluster operators $C(\mathbf{k},s) \to C(\mathbf{k}/s,s)$ (a rescaling of each cluster's momentum by its own size), then a rescaling of all $\mathbf{k}$ by $s$, the total size being integrated over in all terms.
Because the original position variables are defined on a lattice of spacing $a/s$, the momenta are naturally defined in a Brillouin zone of linear size $2\pi s/a$, so this rescaling separates $\mathbf{k}$ from $s$---the size of the Brillouin zone is independent of $s$, and both the function $\phi$ and all cluster operators are evaluated at the same momentum, independent of all cluster sizes.
This simplification is especially useful in the cubic interaction terms. 
We have also extended the size variables down to $0$, and only use an explicit cutoff to some finite size $s_0>0$ when necessary to avoid unphysical divergences stemming from this approximation. 
The interaction constants are labeled to indicate the parity of operators they appear with.
Exchange statistics in the three-cluster interactions requires that
\begin{equation}
    g^{(b;ff)}_{\lambda;\lambda_1\lambda_2} = -g^{(b;ff)}_{\lambda;\lambda_2\lambda_1} \quad \mathrm{and} \quad  g^{(b;bb)}_{\lambda;\lambda_1\lambda_2} = g^{(b;bb)}_{\lambda;\lambda_2\lambda_1}.
\end{equation}
In \cref{eq:gdef} these conditions naturally arise from the (anti)symmetry of the underlying wave functions.
The factors of $1/2$ in the last two terms of the Hamiltonian account for double counting from these symmetries. 

The averages of the three-cluster coefficients are
\begin{equation} \label{eq:g3avgs}
\begin{gathered}
    \expval{g^{(f;fb)}_{\lambda;\lambda_1\lambda_2}} = \expval{g^{(b;ff)}_{\lambda;\lambda_1\lambda_2}} = \expval{g^{(b;bb)}_{\lambda;\lambda_1\lambda_2}} = 0 \\
    \expval{g^{(f;fb)\dagger}_{\lambda;\lambda_1\lambda_2}\, g^{(f;fb)}_{\lambda';\lambda'_1\lambda'_2}} = \frac{g_3^2}{N_\lambda^2}\delta_{\lambda\lambda'}\delta_{\lambda_1\lambda'_1}\delta_{\lambda_2\lambda'_2} \\
    \expval{g^{(b;\eta\eta)\dagger}_{\lambda;\lambda_1\lambda_2}\, g^{(b;\eta\eta)}_{\lambda';\lambda'_1\lambda'_2}} = \frac{g_3^2}{N_\lambda^2}\delta_{\lambda\lambda'}\left(\delta_{\lambda_1\lambda'_1}\delta_{\lambda_2\lambda'_2}+\eta \delta_{\lambda_1\lambda'_2}\delta_{\lambda_2\lambda'_1}\right), 
\end{gathered}
\end{equation}
where $\eta = f$ or $b$, and are all characterized by the same $g_3^2$.
The only difference between the three corresponding terms in the Hamiltonian is the presence or absence of an extra electron inside one of the initial or final clusters, which is insignificant for large clusters.

With all of our approximations the term of \cref{eq:approxH} giving the internal cluster energy from spin-up electrons can be expressed in terms of the density of clusters,
\begin{multline} \label{eq:Einternal}
    E_\uparrow = \int\dd s \sum_{\mathbf{k}} \sum_{\lambda=1}^{N_\lambda}s\,E_p\,\expval{C^\dagger_\lambda(\mathbf{k},s) C_{\lambda}(\mathbf{k},s)} \\
    = E_p \int\dd s\,s\, n_s \propto P\,E_p \propto (1+p)\,E_p
\end{multline}
for $p<p_c$, where we recall $P = (1+p)/2$ is the probability a site belongs to any cluster.
For any given $p$ this and the term proportional to $\mu_p$ are simply constant shifts of the total energy and we drop both from further consideration.

\subsection{Peierls Substitution} \label{sec:peierls}
\vspace{-12pt}
To calculate transport properties we must determine how the to couple the model to an external electromagnetic field. 
Electrons in the original lattice model can be coupled to an electromagnetic field with a Peierls substitution,
\begin{equation}
    t \to t\,\exp[ie\int_{\mathbf{r}}^{\mathbf{r}+\mathbf{e}}\dd\mathbf{l}\cdot\mathbf{A}(\mathbf{l})],
\end{equation}
where $\mathbf{r}$ is the initial position of the electron, $\mathbf{e}$ is the vector giving the direction it hops, $e$ is the electron charge, and $\mathbf{A}$ is the vector potential. 
We will take a spatially uniform  but time dependent $\mathbf{A}$ producing a uniform electric field. 
We then carry through the same approximations as above with this new phase factor.
Just as we discard the internal energy term, we also neglect the coupling of $\mathbf{A}$ to the electrons within clusters.
The terms we keep reflect the coupling of the electromagnetic field to clusters' dynamics rather than just their internal states and allow us to calculate the clusters' electromagnetic properties. 

The result is that the electromagnetic field enters via minimal coupling $\phi(\mathbf{k}) \to \phi(\mathbf{k}-e\mathbf{A})$, and expansion in powers of $e$ lets us treat its effect perturbatively.
To $O(e^2)$ the full Hamiltonian including the coupling to $\mathbf{A}$ can thus be obtained from \cref{eq:approxH} by putting
\begin{equation} \label{eq:Acoupling}
    \phi(\mathbf{k}) \to \phi(\mathbf{k}) - e\,A_i \pdv{\phi(\mathbf{k})}{k_i} + \frac{e^2}{2} A_i A_j \pdv{\phi(\mathbf{k})}{k_i}{k_j},
\end{equation}
where repeated indices are summed. 
Because the function $\phi(\mathbf{k})$ appears with each interaction we acquire two new terms for each cluster interaction, a paramagnetic term at $O(e)$ and a diamagnetic term at $O(e^2)$.
\vspace{-20pt}
\section{Model Action} \label{sec:modelaction}
\vspace{-12pt}
\begin{figure*}[!ht]
    \centering
    \includegraphics[width=0.7\textwidth]{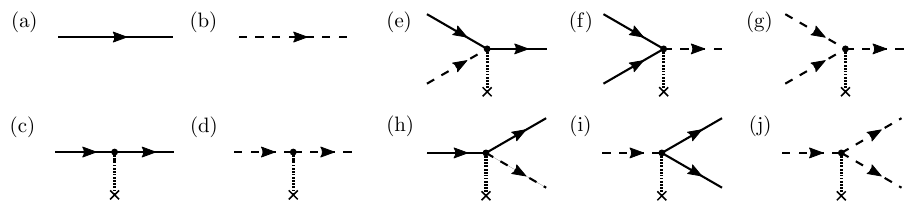}
    \vspace{-10pt}
    \caption{The diagrammatic representation of the terms of our model action, \cref{eq:modelAction}.
    The solid line (a) is the bare fermionic propagator and dashed line (b) is bare bosonic propagator.
    The random Gaussian terms are given in panel (c) for fermions and panel (d) for bosons.  
    The 3-body interactions and their Hermitian conjugates are given in panels (e) through (j). 
    Each $\bullet$ vertex contributes a factor of $t\,\phi(\mathbf{k})$ and the corresponding $g$. 
    Dotted lines in the interaction diagrams terminating in $\times$ represent ``disorder lines'' for each vertex, and connecting two of these gives the average of a square of the corresponding $g$.}
    \label{fig:diagramrules}
\end{figure*}

Using the approximated form of the Hamiltonian \cref{eq:approxH}, we now write a model Matsubara action.
We promote the operators to fields $\psi^\eta_\lambda(\tau,\mathbf{k},s)$ as functions of imaginary time $\tau$, labeling bosonic and fermionic cluster fields by their respective statistical signs $\eta=\pm$, and introduce the appropriate imaginary-time derivative term.
The resulting action is
\begin{widetext}
\begin{align} \label{eq:modelAction}
    S &\approx \int_{0^+}^{s_\xi} \!\!\dd s\! \int_0^{1/T}\!\!\dd\tau \!\int\frac{\dd^2k}{(2\pi)^2} \sum_{\lambda,\lambda'=1}^{N_\lambda} \sum_{\eta=\pm} \bar{\psi}^\eta_\lambda(\tau,\mathbf{k},s)\left[\left(\partial_\tau - \mu_\eta(s)\right)\frac{\delta_{\lambda\lambda'}}{N_\lambda} + t\,\phi(\mathbf{k})\,g_{\lambda\lambda'} \right] \psi^\eta_{\lambda'}(\tau,\mathbf{k},s) \nonumber\\
    &+\int_{0^+}^{s_\xi} \!\!\dd s \! \int_{0^+}^{s} \!\!\dd s'\! \int_0^{1/T}\!\!\dd\tau \!\int\frac{\dd^2 k}{(2\pi)^2} \sum_{\lambda,\lambda_1,\lambda_2}^{N_\lambda}t\,\phi(\mathbf{k})\Bigg[ g^{(f;fb)}_{\lambda;\lambda_1,\lambda_2} \, \bar{\psi}^-_\lambda(\tau,\mathbf{k},s) \psi^-_{\lambda_1}(\tau,\mathbf{k},s-s') \psi^+_{\lambda_2}(\tau, \mathbf{k},s') \nonumber\\
    & + \frac{1}{2}g^{(b;ff)}_{\lambda;\lambda_1,\lambda_2} \, \bar{\psi}^+_\lambda(\tau,\mathbf{k},s) \psi^-_{\lambda_1}(\tau,\mathbf{k},s-s') \psi^-_{\lambda_2}(\tau,\mathbf{k},s') + \frac{1}{2}g^{(b;bb)}_{\lambda;\lambda_1,\lambda_2} \, \bar{\psi}^+_\lambda(\tau,\mathbf{k},s) \psi^+_{\lambda_1}(\tau,\mathbf{k},s-s') \psi^+_{\lambda_2}(\tau,\mathbf{k},s')\Bigg] + \mathrm{h.c.},
\end{align}
\end{widetext}
where $\mu_\pm(s)$ are chemical potentials (distinct from $\mu_p$ in \cref{eq:HubbardH}) introduced to fix the total densities to the cluster number scaling form, $n(s) = n_0 s^{-\tau_P}$ with $n_0$ a proportionality constant
~\footnote{Fixing densities to this form for all $s$ is consistent with viewing this action as an effective IR theory capturing the critical behavior for large $s$. The non-singular contributions from small $s$ in sums involving the full cluster density will be noted as necessary.}.
We can express the terms of this action diagrammatically as shown in \cref{fig:diagramrules}.
The Gaussian propagators are represented as single solid or dashed lines for fermions and bosons respectively. 
With the analogy between our interactions and the random disordered interactions of large-$N$ models, each interaction is associated with a dotted ``disorder'' line ending in $\times$, and sensible diagrams with nonzero averages are made by connecting two $\times$'s from the same type of vertex, representing an average over the square of one of the $g$'s.
\vspace{-24pt}
\subsection{Dyson equation} \label{sec:DysonEq}
\vspace{-12pt}
To calculate properties of this model we first need the Green's functions for fermionic and bosonic cluster fields. 
To proceed we transform to Matsubara frequencies $\omega_n = [2n+(1\mp 1)/2]\pi T$, and identify the noninteracting Green's functions
\begin{equation}
    G_{0,\pm}(i\omega_n,s) = \frac{1}{i\omega_n + \mu_\pm(s)}.
\end{equation}
These are independent of momentum and flavor index, and depend on $s$ only through the chemical potentials. 

\begin{figure*}[!ht]
    \centering
    \includegraphics{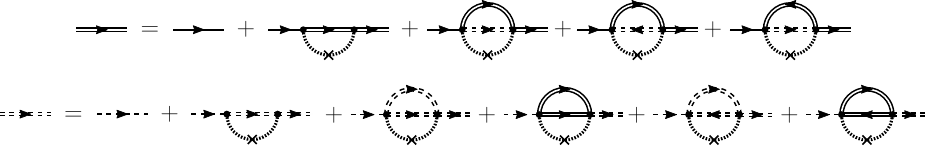}
    \caption{The diagrammatic representation of the full Green's functions, represented here as double solid (fermionic) or double dashed (bosonic) lines.
    Other elements are the same as in \cref{fig:diagramrules}.
    From these diagrams we can write the Dyson equation \cref{eq:Dyson} with self-energies as in \cref{eq:Sigma-,eq:Sigma+}.}
    \label{fig:Dyson}
\end{figure*}

As for other large-$N$ models, interaction contributions are dominated by terms that do not have crossed disorder lines in their diagrammatic representation, for example melon diagrams in the SYK model.
The diagrammatic expansion of the Green's functions for our model are shown in \cref{fig:Dyson} and give the Dyson equation
\begin{equation} \label{eq:Dyson}
    G_\pm(i\omega_n,\mathbf{k},s)^{-1} = i\omega_n + \mu_\pm(s) - \Sigma_\pm(i\omega_n,\mathbf{k},s),
\end{equation}
where the self-energies $\Sigma_\pm$ are most easily written in terms of imaginary time as
\begin{widetext}
\begin{align}
    \Sigma_-(\tau,\mathbf{k},s) &= g_2^2 t^2\phi(\mathbf{k})^2 G_-(\tau,\mathbf{k},s) \nonumber \\*
    & - g_3^2t^2\phi(\mathbf{k})^2 \left[\int_{0^+}^s \!\!\dd s' G_+(\tau,\mathbf{k},s') G_-(\tau,\mathbf{k},s-s') + \int_{0^+}^{s_\xi-s} \!\! \dd s' \sum_{\eta=\pm} \eta\, G_\eta(-\tau,\mathbf{k},s') G_{-\eta}(\tau,\mathbf{k},s+s')\right], \label{eq:Sigma-} \\
    \Sigma_+(\tau,\mathbf{k},s) &= g_2^2 t^2\phi(\mathbf{k})^2 G_+(\tau,\mathbf{k},s) \nonumber \\*
    & - g_3^2t^2\phi(\mathbf{k})^2\! \left[\frac{1}{2}\int_{0^+}^{s} \!\! \dd s'\! \sum_{\eta=\pm} G_\eta(\tau,\mathbf{k},s') G_\eta(\tau,\mathbf{k},s-s') \!+\! \int_{0^+}^{s_\xi-s} \!\! \dd s' \sum_{\eta=\pm} \!\eta\, G_\eta(-\tau,\mathbf{k},s') G_\eta(\tau,\mathbf{k},s+s')\right]. \label{eq:Sigma+}
\end{align}
\end{widetext}
Explicit dependence on $s$ is acquired in the $g_3^2$ terms from the integrals over $s'$. 

We now consider two qualitatively distinct parameter regimes for this set of equations: one where the terms with coefficient $g_2^2$ dominate the self-energy (Gaussian regime), and one where the terms with coefficient $g_3^2$ dominate (cubic regime).
We label quantities in the first case with a subscript $2$ and in the latter case with a subscript $3$, so that
\begin{gather} 
    G_{2,\pm}(i\omega_n,\mathbf{k},s)^{-1} = i\omega_n + \mu_\pm(s) - \Sigma_{2,\pm}(i\omega_n,\mathbf{k},s) \label{eq:G2eqns1}\\
    \Sigma_{2,\pm}(i\omega_n,\mathbf{k},s) = g_2^2 t^2\,\phi(\mathbf{k})^2 G_{2,\pm}(i\omega_n,\mathbf{k},s), \label{eq:G2eqns2}
\end{gather}
and similarly for $G_{3,\pm}$ with $\Sigma_{3,\pm}$ given by the $g_3^2$ terms in \cref{eq:Sigma-,eq:Sigma+}. 
\vspace{-24pt}
\subsection{Gaussian solution \texorpdfstring{--- $G_2$}{}} \label{sec:GaussianG}
\vspace{-12pt}
At the Gaussian level the bosonic and fermionic sectors of the theory have the same form and do not couple so $G_{2,+}$ and $G_{2,-}$ must also have the same form. 
Because $\Sigma_{2,\pm}$ is proportional to $G_{2,\pm}$, we have a quadratic equation for the Green's function and the solution is easily obtained.
Continued from Matsubara frequency to generic complex frequency $z$ we have
\begin{equation}
    G_{2,\pm}(z-\mu_\pm(s),\mathbf{k},s) = \frac{2}{z + \sgn\left[\Re(z)\right] \sqrt{z^2 - 4 g_2^2 t^2\,\phi(\mathbf{k})^2}},
\end{equation}
where the sign before the square root ensures that $G_2(z) \sim 1/\abs{z}$ for large $\abs{z}$. 

\begin{figure}[b]
    \centering
    \includegraphics[width=\columnwidth]{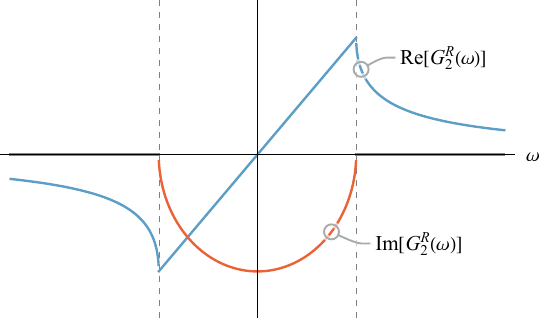}
    \caption{A plot of the real and imaginary parts of the retarded Gaussian sector Green's functions $G^R_{2,\pm}$, \cref{eq:G2R}.
    The vertical dashed lines indicate where $\abs{\omega+\mu_\pm} = 2g_2t\abs{\phi(\mathbf{k})}$.}
    \label{fig:G2plot}
\end{figure}

Restricting to $z=\omega+i0$ gives the retarded Green's functions,
\begin{multline} \label{eq:G2R}
    G^R_{2,\pm}(\omega-\mu_\pm(s),\mathbf{k},s) = \frac{\Theta(\omega^2 - 4g_2^2 t^2\,\phi(\mathbf{k})^2)}{\frac{\omega}{2} + \sgn(\omega)\sqrt{\left(\frac{\omega}{2}\right)^2 - g_2^2 t^2\,\phi(\mathbf{k})^2}} \\
    + \left(\frac{\omega}{2} - i \sqrt{g_2^2 t^2\,\phi(\mathbf{k})^2-\left(\frac{\omega}{2}\right)^2}\right) \frac{\Theta(4g_2^2 t^2\, \phi(\mathbf{k})^2 - \omega^2)}{g_2^2 t^2\,\phi(\mathbf{k})^2}
\end{multline}
shown in \cref{fig:G2plot}, and the spectral densities are
\begin{multline}
    \rho_{2,\pm}(\omega,\mathbf{k}) = -\frac{1}{\pi}\Im G^R_{2,\pm}(\omega-\mu_\pm(s),\mathbf{k},s) \\
    = \frac{\sqrt{g_2^2 t^2\,\phi(\mathbf{k})^2-\left(\frac{\omega}{2}\right)^2}}{\pi\, g_2^2 t^2\,\phi(\mathbf{k})^2} \Theta\left(4g_2^2 t^2\,\phi(\mathbf{k})^2-\omega^2\right).
\end{multline}
Notice that the range of support for the spectral density depends on $\phi(\mathbf{k})^2$, which is largest for small $ka$ where this IR effective theory applies. 

Using these spectral densities we write the bosonic and fermionic cluster densities,
\begin{equation}
    n_\pm(s) = \int \frac{\dd^2 k}{(2\pi)^2} \int \dd\omega\, \rho_{2,\pm}(\omega,\mathbf{k})\, n_{B/F}(\omega-\mu_a(s)),
\end{equation}
where $n_{B/F}(\omega) = 1/(e^{\omega/T}\mp 1)$ are the Bose and Fermi functions.
We fix the chemical potentials by ensuring these densities are consistent with the behavior of the percolation scaling theory; since $n(s)$ is small, both $n_+(s)$ and $n_-(s)$ must also be small.
Therefore, for $T\ll t$ momenta satisfying $\abs{ka}\ll 1$ are most relevant, the values for which $\rho_{2,\pm}(\omega,\mathbf{k})$ has support at the lowest frequencies. 
For large enough $T$ we find
\begin{multline} \label{eq:mu}
    \mu_\pm(s) \approx -2T\arcoth\left[\left(1\pm \frac{2n_\pm(s)}{n_\Lambda}\right)^{\pm1}\right] \\
    \approx T\ln\left(\frac{n_\pm(s)}{n_\Lambda}\right)
\end{multline}
where $n_\Lambda$ is the density obtained by filling states up to energy at which we must cut off the small $\abs{ka}$ approximation and is much larger than $n_\pm(s)$, which allows the final approximation in terms of a log. 

\subsection{Cubic solution \texorpdfstring{--- $G_3$}{}} \label{sec:CubicG}
\vspace{-12pt}
Our analysis in the cubic regime is similar to the treatment of the SYK model for complex fermions, e.g. as in Refs.~\cite{Sachdev2023,Gu2020}. 
(For details of this full calculation see \cref{sec:G3details}).
We first assume that the self-energies $\Sigma_{3,\pm}$ dominate over the linear-in-frequency term of the Dyson equation, which will partially restrict the regime in which the resulting solution applies.  
Then the form of the self-energies motivates us to consider a scaling ansatz for $G_{3,\pm}$ in the zero temperature limit, and for complex frequency $z$ (with $\Im(z)>0$) we put
\begin{equation}
    G_{3,\pm}(z,\mathbf{k},s) = s^{\gamma_\pm} \frac{e^{-i(\pi\Delta_\pm + \theta_\pm(s))}}{\Omega_\pm(\mathbf{k})^{2\Delta_\pm} z^{1-2\Delta_\pm}},
\end{equation}
where $\gamma_\pm$ and $\Delta_\pm$ are real exponents characterizing the scaling with $s$ and frequency respectively, $\Omega_\pm(\mathbf{k})$ are energies introduced to fix dimensions and account for momentum dependence, and $\theta_\pm(s)$ parametrize the spectral asymmetry and consequently the finite density for each $s$---they are in principle related to $\mu_\pm(s)$ which fix the density in the nonscaling regime. 

Positivity of the spectral densities gives the constraints $\pi\Delta_+ \leq \theta_+(s) \leq \pi(1-\Delta_+)$ and $-\pi\Delta_- \leq \theta_-(s) \leq \pi\Delta_-$. 
Furthermore, since the density $n(s)$ vanishes for $s>s_\xi$, we take the functions $\theta_\pm(s)$ to depend on $s_\xi$ itself and on $s$ only through $s/s_\xi$, and we substitute $\theta_\pm(s) \to \theta_\pm(s/s_\xi)$. 
Transforming to imaginary time $\tau$ the ansatz becomes
\begin{multline}
    G_{3,\pm}(\tau,\mathbf{k},s) = -\sgn(\tau) \frac{s^{\gamma_\pm}\Gamma(2\Delta_\pm)}{\pi\Omega_\pm(\mathbf{k})^{2\Delta_\pm} \abs{\tau}^{2\Delta_\pm}}\\
    \times \sin\left[\pi\Delta_\pm + \sgn(\tau)\theta_\pm(s/s_\xi)\right],
\end{multline}
where $\Gamma(x)$ is the gamma function. 

We assume that the self-energies evaluated at zero frequency cancel against the chemical potentials in \cref{eq:Dyson}, $\Sigma_\pm(0,\mathbf{k},s) = \mu_\pm(s)$. 
Then substituting the ansatz we find that consistency of the $s\to0^+$ and $s=s_\xi$ limits of the Dyson equation constrain many of our parameters: we obtain exponents $\Delta_+ = \Delta_- = 1/3$ and $\gamma_+ = \gamma_- = -1/3$, and find $\Omega_+(\mathbf{k}) \propto \Omega_-(\mathbf{k}) \propto t\abs{\phi(\mathbf{k})}$.
The constants $c_\pm$ that can be introduced to exactly relate $\Omega_\pm(\mathbf{k})$ to $t\,\phi(\mathbf{k})$ are $O(1)$ numbers given by integrals over functions of $\theta_\pm(s/s_\xi)$.

Altogether we have
\begin{equation} \label{eq:G3}
    G_{3,\pm}(z,\mathbf{k},s) = \frac{e^{-i(\pi/3 + \theta_\pm(s/s_\xi))}}{c_\pm^{2/3}g_3^{2/3}s^{1/3}t^{2/3}\phi(\mathbf{k})^{2/3}z^{1/3}},
\end{equation}
the same form for both bosonic and fermionic fields. 
Taking $z\to\omega+i0$ we acquire the retarded Green's function,
\begin{equation}
    G^R_{3,\pm}(\omega,\mathbf{k},s) = \sgn(\omega)\frac{e^{-i(\sgn(\omega)\pi/3+\theta_\pm(s/s_\xi))}}{c_\pm^{2/3}g_3^{2/3}s^{1/3}t^{2/3}\phi(\mathbf{k})^{2/3}\abs{\omega}^{1/3}}.
\end{equation}

To obtain this solution we have assumed first that we are in a parameter regime where the Gaussian terms in the self-energies are much smaller than the cubic terms, and second that $\abs{\Sigma_{3,\pm}} = \abs{G_{3,\pm}^{-1}} \gg \abs{\omega}$ for nonzero frequency.
These constrain the frequency regime for which $G_{3,\pm}$ applies to be
\begin{equation} \label{eq:G3R}
    \frac{g_2^3}{s\,g_3^2}t\abs{\phi(\mathbf{k})} \ll \abs{\omega} \ll \sqrt{s}\,g_3\,t\abs{\phi(\mathbf{k})}.
\end{equation}
Since we are interested large $s$ we therefore find that this power-law Green's function may apply over a finite range of frequencies, not just in the limit of zero frequency as is the typical case for the SYK model---indeed, this solution \emph{never} applies for zero frequency, and in that limit we must use $G_2$ as found in \cref{sec:GaussianG}; $s_\xi$ is always cut off at some large value in a finite system, so the regime of applicability for $G_3$ never extends down to include $\omega = 0$.

\section{Transport properties} \label{sec:transport}
\begin{figure}[b]
\vspace{3mm}
    \centering
    \includegraphics[width=\columnwidth]{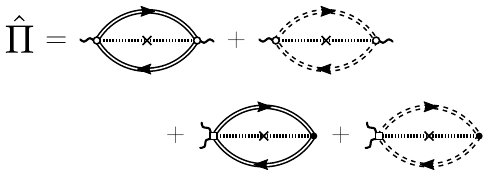}
    \caption{The diagrams giving the response of the effective theory to an external electromagnetic field, dropping all contributions $O(1/N_\lambda)$ or smaller.
    The open circles with a single $\sim$ represent paramagnetic coupling to $\mathbf{A}$ and the squares with two $\sim$'s represent diamagnetic coupling to $\mathbf{A}^2$.
    The propagators (double solid or dashed lines) are dressed by interactions as in \cref{fig:Dyson}.
    All other diagrams vanish with the disorder average or have crossed impurity lines and so vanish in the $N_\lambda\to\infty$ limit.}
    \label{fig:response}
\end{figure}

\vspace{-12pt}
The coupling of the cluster fields to a spatially-uniform electromagnetic field in the action \cref{eq:modelAction} is obtained via minimal coupling as in \cref{sec:peierls} for the Hamiltonian description, but now we take the vector potential to be a function of imaginary time $\mathbf{A}(\tau)$. 
The electromagnetic linear response function $\hat{\Pi}(i\omega_n)$ can be calculated in the usual way as the second derivative with respect to $\mathbf{A}$ of the partition function constructed from the action $S$ including the coupling to $\mathbf{A}$. 
The result is greatly simplified because of the large-$N$ nature of the theory and the averaging procedure.
First, there are no contributions at first order in an interaction constant since $\expval{g_{\dots}}=0$. 
Therefore, the two types of diagrams contributing to $\hat{\Pi}$ are bubble diagrams with two paramagnetic vertices or with one diamagnetic vertex and one normal interaction vertex. 
Second, vertex corrections necessitate crossed disorder lines, which contribute at $O(1/N_\lambda)$, and so vanish in the large-$N_\lambda$ limit.
Finally, because $g_2^2 \gg g_3^2$ the primary contribution comes from the coupling of the Gaussian terms in the action to $\mathbf{A}$, so $\hat{\Pi}$ has an overall factor of $g_2^2$. 
Taking these points into account the effect of diagrams higher order in $g_2$ and $g_3$ is simply to dress the bare propagators as in \cref{fig:Dyson}, so $\hat\Pi$ can be represented diagrammatically as in \cref{fig:response}, and we have
\begin{multline} \label{eq:Pi}
    \Pi_{ij}(i\omega_n) \approx e^2 t^2 g_2^2\, T \int\! \dd s \int\frac{\dd^2k}{(2\pi)^2} \sum_{\eta=\pm} \sum_{\{\omega_n'\}} \\
    \times \left(\pdv{\phi(\mathbf{k})}{k_i}\pdv{\phi(\mathbf{k})}{k_j} G_\eta(i\omega_n',\mathbf{k},s) G_\eta(i\omega_n'-i\omega_n,\mathbf{k},s) \right.\\
    \left.+\phi(\mathbf{k})\pdv{\phi(\mathbf{k})}{k_i}{k_j} G_\eta(i\omega_n',\mathbf{k},s)^2\right).
\end{multline}
The optical conductivity is defined using the retarded response, obtained from this via analytic continuation $i\omega_n \to \omega+i0$.
Since the model is isotropic and we are interested in its longitudinal transport properties, we further define the longitudinal response function as the average of the diagonal elements of this matrix, $\Pi = (\Pi_{xx}+\Pi_{yy})/2$.

Even though the Gaussian couplings to $\mathbf{A}$ dominate $\Pi$, the Green's functions appearing in \cref{eq:Pi} may have either Gaussian $G_2$ or cubic $G_3$ form; whether the Gaussian or cubic self-energies dominate the Dyson equation depends on the frequency regime being considered, not the relative sizes of $g_2$ and $g_3$. 
When evaluating the Matsubara sum in the standard way via contour integration the different analytic structures of $G_2$ and $G_3$ become vitally important and each case must be considered separately.
As previously noted, when continued to real frequencies $G_3$ is applicable in a range of finite frequencies excluding $\omega=0$, so that $G_2$ is relevant for the dc limit. 

\subsection{Linear-in-T Resistivity}
\vspace{-12pt}
The dc resistivity $\rho$ is the reciprocal of the dc conductivity $\sigma$, which itself is the real part of the $\omega\to0$ limit of the optical conductivity.
In terms of $\Pi$,
\begin{equation}
    \sigma = \Re\left[\lim_{\omega\to0} \sigma(\omega)\right] = \Re\left[\lim_{\omega\to0} \frac{\Pi(\omega+i0)}{i\omega}\right]. 
\end{equation}
After rewriting the Matsubara sums in the expression for $\Pi$ in terms of integrals over real frequencies of Fermi and Bose distribution functions and functions of retarded and advanced Green's functions we have
\begin{widetext}
\begin{multline} \label{eq:sigma}
    \sigma = \Re\left\{-\frac{e^2 t^2 g_2^2}{4\pi} \int \dd s \int\frac{\dd^2k}{(2\pi)^2} \abs{\grad_k \phi(\mathbf{k})}^2 \sum_{\eta=\pm} \int\dd\omega' n_a(\omega') \lim_{\omega\to0} \frac{1}{\omega}\left[G^R_{\eta}(\omega') - G^A_{\eta}(\omega') \right] \left[G^R_{\eta}(\omega'+\omega) + G^A_{\eta}(\omega'-\omega) \right]\right. \\
    \left.- \lim_{\omega\to0} \frac{1}{\omega} \frac{e^2 t^2 g_2^2}{4\pi} \int \dd s \int\frac{\dd^2k}{(2\pi)^2} \phi(\mathbf{k})\grad_k^2 \phi(\mathbf{k}) \sum_{\eta=\pm} \int\dd\omega' n_a(\omega') \left[G^R_{\eta}(\omega')^2 - G^A_{\eta}(\omega')^2 \right]\right\},
\end{multline}
\end{widetext}
where we have suppressed the Green's functions' common dependence on $\mathbf{k}$ and $s$. 
In this regime the relevant Green's functions are the zero-frequency limit of $G_{2,\pm}$ as in \cref{eq:G2R} with the chemical potentials $\mu_\pm(s,T)$ as in \cref{eq:mu}.
Because of the branch cut in these Green's functions for small frequencies, the integral over the integrated frequency $\omega'$ is restricted to the range $-2g_2t\abs{\phi(\mathbf{k})}-\mu_\pm(s) \leq \omega' \leq 2g_2t\abs{\phi(\mathbf{k})} - \mu_\pm(s)$;
contributions from outside this range exactly cancel.
Because of the $\omega\to0$ limit we expand the Green's functions in the first line in powers of $\omega$.
The zeroth-order terms of this expansion, which simply evaluates the Green's functions at $\omega=0$, are overall proportional to $1/\omega$ and combine with the second line.
Using the form of $G^R_{2,\pm}$ and $G^A_{2,\pm} = (G^R_{2,\pm})^*$, these terms are found to contribute a purely imaginary and divergent term $\propto 1/(i\omega)$ as expected from a model with translational symmetry.
In a clean electron gas a $1/(i\omega)$ divergent Drude term captures the unimpeded acceleration of charged quasiparticles in the presence of an electric field, and is regulated by disorder by introducing a quasiparticle scattering rate.
Here, however, there are no coherent quasiparticles so this understanding of the meaning of this divergence and how it is regulated by disorder does not obviously apply.
We will not analyze here how or if this term is regulated or removed by the inclusion of translation symmetry breaking disorder in the system, as it must be since infinite conductivity is unphysical.
We will drop this term from primary consideration for now. 
Of the remaining terms in the expansion of the Green's functions only the terms first order in $\omega$ survive in the $\omega\to0$ limit, and will be our focus from here on.

The Green's functions in the remaining real term enter in the form
\begin{multline}
    \left[G^R_{\pm,2}(\omega') - G^A_{\pm,2}(\omega') \right] \left[\dv{G^R_{\pm,2}(\omega')}{\omega'} - \dv{G^A_{\pm,2}(\omega')}{\omega'} \right] \\
    = \frac{1}{2}\dv{\omega'}\left[G^R_{\pm,2}(\omega') - G^A_{\pm,2}(\omega')\right]^2 = \frac{\omega'+\mu_\pm(s)}{g_2^4 t^4\phi(\mathbf{k})^4},
\end{multline}
where the last equality uses the form of the Green's function in this frequency range, \cref{eq:G2R}. 
With a shift of the integrated frequency by the chemical potential the dc conductivity becomes
\begin{multline} \label{eq:sigmadc1}
    \sigma = -\frac{e^2}{4\pi t^2 g_2^2} \int\dd s\int\frac{\dd^2k}{(2\pi)^2}\frac{\abs{\grad_k \phi(\mathbf{k})}^2}{\phi(\mathbf{k})^4}\\
    \times \sum_{\eta=\pm} \int_{-2g_2t\abs{\phi(\mathbf{k})}}^{2g_2t\abs{\phi(\mathbf{k})}} \dd\omega'\, \omega' \, n_\eta\left[\omega'-\mu_\eta(s)\right],
\end{multline}
and we can evaluate the frequency integral using the chemical potentials given in \cref{eq:mu}.
If we take $\abs{\mu_\pm(s)} \gg 2g_2t\abs{\phi(\mathbf{k})}$, equivalent to  $T/t \gg g_2/\abs{\log(n_\pm(s)/n_\Lambda)}$, then all frequencies in the integration region are small compared to the chemical potential and we can expand the distribution function in \cref{eq:sigmadc1} in powers of $\omega'$.
The lowest order term gives
\begin{multline}
    \sigma \approx -\frac{4e^2 g_2t}{3\pi} \int\dd s \int\frac{\dd^2k}{(2\pi)^2}\frac{\abs{\grad_k \phi(\mathbf{k})}^2}{\abs{\phi(\mathbf{k})}} \sum_{\eta=\pm} n'_\eta\left(\abs{\mu_\eta(s)}\right) \\
    \propto \frac{e^2 g_2}{m\, T} n_C(p)
\end{multline}
with the higher order terms contributing with higher powers in the small constant $g_2$, and so can be neglected.
To obtain this result we use $n_\pm(s) \ll n_\Lambda$ to expand the derivatives of the Bose and Fermi functions and replace $n_+(s)+n_-(s) = n(s)$, then write the integral of this function over $s$ as the total cluster density $n_C$ as in \cref{eq:clusterdensity}. 
We use the $\abs{ka}\ll 1$ expansion of $\phi(\mathbf{k})$ to perform the momentum integral, and define the effective mass as $m = 1/(2ta^2)$. 
The resistivity is the inverse of this expression, and keeping just temperature and doping dependence we obtain
\begin{equation}
    \rho \sim \frac{T}{n_C(p)}.
\end{equation}
Though the precise dependence on $p$ here is determined by the values of the expansion coefficients in \cref{eq:clusterdensity}, $\rho$ takes its minimum value at $p=p_c$---the proliferation of large finite clusters near the critical point should yield the minimum cluster density and therefore maximum resistivity. 
Though we explicitly dropped the imaginary divergent part of $\sigma(\omega\to0)$ since the conductivity of a physical system cannot actually be divergent, if analyzed in this same way we find that it acquires exactly the same dependence on $T$ and $p$ as the real part we have analyzed here explicitly.
If breaking translation symmetry introduces a simple constant low-frequency cutoff in this term, then its inclusion into $\rho$ would not affect either the $T$ or $p$ dependence.
However, how this might be cut off at low frequencies is beyond the scope of the model as currently formulated. 

Curiously, we have found a linear-in-$T$ resistivity stemming from the Gaussian sector of the model action \cref{eq:modelAction}, which is similar in the form to a random matrix model.
The random matrix model has a description in terms of quasiparticles---$g_{\lambda\lambda'}$ can be immediately diagonalized instead of following a Green's function calculation as in \cref{sec:GaussianG}---and therefore is known not to give such anomalous behavior~\cite{Sachdev2023}.
Here, however, the parameter $s$ and appearance of the momentum-dependent function $\phi(\mathbf{k})$ in the kinetic energy distinguish this model from a random matrix model and allows this result to be obtained.
The only energy scales appearing in the spectral densities are the frequency $\omega$ and the kinetic energy $g_2t\abs{\phi(\mathbf{k})}$, both of which are integrated to find the cluster field densities and set the chemical potentials. 
Thus, in the parameter regime considered the only energy or length scales remaining inside $\mu_\pm(s)$ are the cluster densities $n(s)$ and the temperature. 
Since $s$ is also integrated, the only scale remaining in the end to determine the behavior of $\rho$ is $T$. 

The approximations made to obtain this result imply that the linear-in-$T$ behavior does not persist down to $T=0$ or up to $T\to\infty$, but for a nevertheless wide temperature range $g_2/\abs{\log(n(s_0)/n_\Lambda)} \ll T/t \ll 1$.
At this lower limit contributions from higher odd powers of $T$ start to become relevant.
We do not have a numerical estimate for the small constant $g_2$ nor the other factors entering into this lower scale, however, so it is not possible to say precisely at what temperature scale we might expect to see a significant deviation from linear-in-$T$ behavior. 
Regarding the upper limit, if $t\sim 100\mathrm{meV} \sim 10^3\mathrm{K}$ then we expect nonsaturating linear-in-$T$ behavior up to the highest experimentally accessible temperatures in real systems. 

\subsection{Power-law Optical Conductivity}
\vspace{-12pt}
As found in \cref{sec:CubicG}, the cubic Green's functions $G_{3,\pm}$ apply over a wide range of frequencies between UV and IR limits for large $s$. 
Using these Green's functions in $\Pi$ we can therefore obtain the leading-order contributions to the optical conductivity relevant in this intermediate frequency regime.
A simple scaling analysis of the frequency dependence of this optical conductivity gives
\begin{multline}
    \sigma(\omega) = \frac{\Pi(\omega+i0)}{i\omega} \\
    \sim \frac{1}{\omega} \int \dd\omega' \sum_{\eta=\pm} G^R_{3,\eta}(\omega') G^A_{3,\eta}(\omega'-\omega) \\
    \sim \abs{\omega}^{-2/3} e^{i\pi/3}.
\end{multline}
We therefore find that the leading contribution to $\sigma(\omega)$ for this model is of power-law form for a range of intermediate frequencies with exponent $\gamma = 2/3$.
Higher order terms, e.g. from the cubic terms coupling to photons at $O(g_3^2)$, will contribute terms with other frequency dependence but small coefficients, so that $\sigma(\omega)$ will not have a true power-law form but will be well-approximated by a power-law with $\gamma\approx 2/3$.
This is consistent with the results of experiments in the cuprates~\cite{Cooper1990, Schlesinger1990, ElAzrak1994, Baraduc1996, Marel2003, Hwang2007}, where an apparent power-law varies between samples and with doping.

To recover this behavior we only needed a simple scaling argument using the $T=0$ form of the cluster Green's functions.
In principle, a more precise form for the optical conductivity may be obtained from a more complete analysis, extending these Green's functions to nonzero temperature leveraging the cubic regime's imaginary time reparametrization invariance and a conformal transformation, e.g. as discussed in Refs.~\cite{Parcollet1999, Sachdev2015}.
We will leave such analysis to later work. 

\section{Discussion and Outlook} \label{sec:discussion}
\vspace{-12pt}
The simple first analysis of the transport properties of the model in the previous section yields clear non-Fermi-liquid properties. 
The leading contribution to the optical conductivity has power-law form $\sigma(\omega) \sim \abs{\omega}^{-\gamma}$ with $\gamma = 2/3$, and the dc resistivity is found to be linear-in-$T$ with doping-dependent slope that takes its maximum value at a well-defined critical doping $p_c$ depending on the lattice geometry.  
Interestingly, these behaviors are found to arise from different sectors of the model. 
For zero and nonzero frequency, different contributions to the self-energies dominate, so although all terms in the model action result from implementing a single principle---a classical percolation transition---particular anomalous properties characteristic of the strange metal phase do not all arise from the same microscopic processes. 
The dc properties arise from the Gaussian sector of the theory, with the corresponding cluster interaction term related to the shifting of single clusters, while the optical properties arise from the cubic sector, describing the merging and dividing of clusters. 

Here we have focused exclusively on transport in the absence of magnetic fields, but anomalous Hall response and magnetoresistance are also seen experimentally in the same range of dopings. 
Because the construction presented here depends the system possessing spin symmetry while also treating the two electron spins separately---one bounding clusters, the other residing within them---a natural question arises as to how to account for the partial magnetization of the system in response to the external field necessary for these effects.
Our choice here of having spin-down electrons bound clusters was arbitrary, and exchanging spin-up and spin-down everywhere yields the same results.
If the system has an excess of one spin species, however, then this symmetry is broken and the doping yielding the percolation transition would shift up or down from the symmetric case depending on the choice.
Putting $n_\uparrow = (n_e+\delta)/2$ and $n_\downarrow = (n_e-\delta)/2$, then identifying the critical hole doping as a percolation transition as in \cref{sec:HU} gives $p_{c\uparrow} \approx p_c + \delta$ and $p_{c\downarrow} \approx p_c - \delta$ depending on whether spin-up or spin-down electrons are taken to bound clusters, where here $p_c$ is the spin-symmetric result. 
If $\delta$ is smaller than the uncertainty of the doping level in experiments then the effect of this splitting may not have observable consequences, and only subsequent effects of an applied magnetic field in our analysis would be necessary to consider. 
If $\delta$ is not small, however, then a qualitatively new calculation is likely required that can manage the split percolation transition for different spins. 

The effect of disorder, which will be present in any real system, has not been considered here in any detail.
Though we modeled the matrix elements $g$ for the cluster dynamics terms as random quantities, we began from the clean Hubbard Hamiltonian on a perfectly regular lattice. 
If, at the level of the Hubbard Hamiltonian, we were to include quenched hopping disorder or an on-site disorder potential characterized by an energy $W\ll t,U$, then the initial rewriting of the system in terms of clusters is largely unaffected, and our ignoring of rare spatially ordered states would be more explicitly justified. 
Furthermore, the interaction constants $g$ would be affected by either of these forms of disorder to become actually random quantities, now depending explicitly on a spatial coordinate like the model in Ref.~\cite{Patel2023}.

Though the properties we calculate here for the square lattice are intriguingly consistent the phenomenology of the cuprates when taken at face value, the model cannot be directly applied to explain these real materials as currently formulated.
The infinite-$U$ limit we use to obtain the cluster model from the Hubbard model sets the energy $J\sim t^2/U$ characterizing the exchange interaction generated by virtual double occupation of sites to zero.
Therefore, the only potential regime where this theory applies with regards to the cuprates is at temperatures above this scale, far above the regime where the strange metal is defined experimentally.
Though we have occasionally used the term ``strange metallic'' to describe the properties 
To make any concrete claims from this model about real systems, the infinite-$U$ limit must be relaxed to give a finite $J$, and extended to more physically relevant parameter regimes. 
In this context, what is presented here can be viewed as the zeroth order expansion in $J$.

This brings us to a number of general questions about how this theory might be extended to understand the low temperature effects of real systems. 
In the cuprates, for $p<p_c$ strange metallic behavior gives way to the pseudogap phase below a crossover temperature $T^*$ that decreases with increasing hole doping and drops to $0$ at $p=p_c$~\cite{Proust2019}.
Since $T^*$ vanishes at $p_c$, might it be related to some critical behavior of the percolation transition, or is the pseudogap fundamentally connected to the exchange interaction parameterized by $J$ which this theory does not include?
In the former case, the correlation length characterizing the largest cluster $\xi$ provides one natural IR cutoff for the theory, $E_\xi \sim 1/\xi \sim \abs{p-p_c}^{\nu_P}$, which decreases with increasing $p$ and vanishes at the critical point.
On length scales larger than $\xi$ or energy scales smaller than $E_\xi$ the cluster structure cannot be distinguished and the theory investigated here is invalid, which may signal one avenue to approach the emergence of a pseudogap, or at the very least a new phase of this model with distinct properties from what we find here.

Perhaps most importantly, superconductivity arises in many strongly correlated systems, most notably in the cuprates in a low-temperature dome covering a range of dopings around $p_c$. 
The model developed here does not possess a mechanism for forming Cooper pairs, which is naturally provided by the exchange interaction. 
This interaction favors the formation of spin singlet states on nearest neighbor sites, which in the cluster picture occurs at the boundaries of the clusters---spin-down electrons bound the clusters filled with spin-up electrons.
Therefore we see that the sorts of clusters we have been considering, which have fractal properties on average and large surface areas, are energetically favored by including this additional term in the Hamiltonian compared to clusters with a lower ratio of perimeter to enclosed area, for example a more circular geometry.
Additionally, when $T\sim J$ the pairing effect of this interaction will become important and may dominate, causing the principle assumptions of this work---uncorrelated spin-up and spin-down electron positions---to fail. 
It is currently unclear how to include all effects of this $J$ interaction into \cref{eq:modelAction} and therefore we cannot yet explore the possibility of superconductivity in this model, or indeed how this interaction would affect any other low-temperature behaviors.

Turning from temperature to doping, we have focused exclusively on $p\leq p_c$ since above this value an infinite cluster appears which requires qualitatively new considerations.
The model action \cref{eq:modelAction} is nevertheless still relevant for describing the behavior of finite clusters above the critical doping, so the contributions to the transport properties that we have calculated will remain, though now accompanied by contributions from the infinite cluster.
In finite clusters electrons are strictly confined---they may not pass through the large energy barrier at the boundary---but this is not the case for the infinite cluster; 
in principle electronic wave functions are not forbidden from being extended, though localization may also result from interference effects in any given configuration of the system~\cite{Mookerjee1995,Ostrovsky2014}.
However this quantum percolation problem, as it is called, is somewhat different than what would arises in a percolation-based recasting of the Hubbard model for $p>p_c$, since the infinite cluster itself is not static.
The electrons forming its boundary have dynamics themselves, and further study would be needed to determine how this affects the nature of the electronic states within, i.e. whether states are extended, power-law localized, or exponentially localized. 
Whatever the case, the infinite cluster surely provides a crucial contribution to transport properties above $p_c$, especially as it grows with higher hole doping to encompass an ever larger portion of the system---just above the transition the fraction of sites within the infinite cluster is $P_\infty \propto (p-p_c)^{\beta_P}$ with critical exponent $\beta_P = 5/36$ in two dimensions. 
An energy associated with this order parameter $P_\infty$ or some other property of the infinite cluster may plausibly set the temperature scale where a crossover to a Fermi-liquid-like state is observed in the overdoped cuprates. 

Though the cuprates provide an important but imperfect point of comparison for this theory since they realize aspects of the Hubbard model on a square lattice, the conclusions found here apply more generally.
The geometry of the lattice played little role in our analysis apart from determining the value of the critical doping and the precise form of the function $\phi(\mathbf{k})$, so all of the anomalous behaviors we have obtained should also manifest in other lattices near other specific hole dopings---near $p\sim0.39$ for the hexagonal lattice and near $p\sim 0.31$ for the kagome lattice. 
Observing linear-in-$T$ resistivity and power-law optical conductivity in these lattices near these dopings would be a strong indication that the behavior of the square lattice near $p \sim 0.19$ is also captured by this theory, and by extension may be a good starting point for a model capturing aspects of the cuprate normal state. 
Though solid-state systems realizing the Hubbard model on these lattices are not known to exist, to say nothing of these high hole dopings, there is no fundamental obstruction to simulating such systems numerically or with cold atoms in optical lattice experiments, as have been done for the square lattice~\cite{Vidhyadhiraja2009, Huang2019, Wu2022, Xu2019, Brown2019}.
Even including the square lattice, however, there are tests of the theory that can be done.
The doping-dependent transition, key to writing the low energy model, entails a distribution of clusters of sites obeying a critical theory. 
Experiments in cold atoms can directly image the real-space configuration of a Hubbard system, which should then display critical features if this transition is indeed present, e.g. a power-law distribution of cluster sizes $n_s \propto s^{-\tau_P}$ near $p_c$, or a largest cluster size that scales as $s_\xi \propto \abs{p-p_c}^{-1/\sigma_P}$. 

One of the broader goals of this work is to bridge the gap between the Hubbard model, a well accepted microscopic starting point for the analysis of real strongly-correlated systems, and theories without quasiparticles developed to study strange metallic behavior, which often lack a clear connection to the microscopic physics of electrons in a lattice. 
We have shown that utilizing the structure of a classical percolation transition in basis states of the large-$U$ Hubbard model motivates an effective model that exhibits a number of key anomalous behaviors, and has features of a number of different types of theories used to study strange metals: it is a large-$N$ model like SYK and its relatives, the cluster size $s$ plays a similar role to the continuous mass spectrum in theories of unparticles, and the important degrees of freedom are those exhibiting critical properties from a doping-driven transition as in theories with a Fermi surface coupled to critical bosons. 
In this theory the important degrees of freedom originate from extended clusters of electrons in the original real-space electronic basis, and the large number of ``flavors'' in the model are related to the large number of possible shapes for large critical clusters near the percolation transitions. 
Ultimately, the non-Fermi-liquid transport of the model is due the scaling properties of these clusters and the unusual nature of their dynamics. 

The ability of our effective theory \cref{eq:modelAction} to reproduce several notable phenomenological features associated with the cuprate strange metal without finely tuned parameters or \emph{ad hoc} additional degrees of freedom, and to naturally explain why this behavior would manifest near $p\sim 0.19$ specifically, suggests that it may capture relevant physics of the strange metallic phase, despite the limitations inherent in the assumptions made to develop it.
There are clear ways to test the validity of the theory and to continue theoretical analysis, both by deeper investigations of aspects of the theory presented here and by extending it, e.g. including additional weaker hopping terms that could couple electrons in different clusters, or by including the exchange interaction. 
Most importantly, these extensions could provide a starting point for analytical studies of superconductivity and other low-temperature phases of the Hubbard model.

\begin{acknowledgments}
I thank Nigel Cooper, Pieter Claeys, Jan Behrends, Justin Wilson, Zachary Raines, Daniel Sheehy, Ilya Vekhter, Haoyu Guo, and Elio K\"onig for helpful discussions. 
This work was performed in part at the Aspen Center for Physics, which is supported by the National Science Foundation Grant No.~PHY-1607611. 
\end{acknowledgments}

\appendix
\section{Boundary electron hopping matrix elements} \label{app:gcalculation}
\vspace{-12pt}
In an initial cluster set $\mathscr{C}$, there are $N_c$ clusters that have a particular spin-down electron forming part of their boundary, and when this electron hops we are taken to a different cluster set $\mathscr{C}'$ in which those initial clusters have been changed into $N'_c$ new clusters. 
We label the initial clusters $\zeta_1,\dots,\zeta_{N_c}$ and the final clusters $\zeta'_1,\dots,\zeta'_{N'_c}$.
We define the union of the sites in the initial and final sets of clusters respectively as $Z_c$ and $Z'_c$. 
Since the total number of spin-up electrons within the initial and final clusters cannot change we have
\begin{equation}
    \nu = \sum_{i=1}^{N_c} \nu_i = \sum_{j=1}^{N'_c} \nu'_j = \nu',
\end{equation}
and similarly the total number of sites in the initial and final sets of clusters cannot change, so we have
\begin{equation}
    s = \sum_{i=1}^{N_c} s_i = \sum_{j=1}^{N'_c} s'_j = s'. 
\end{equation}
We also know how the total center of mass of the involved clusters moves; the sets $Z_c$ and $Z'_c$ differ in the location of a single site, which is displaced by $-\mathbf{e}$, so the the total final center of mass $\mathbf{r}'$ is related to the total initial center of mass $\mathbf{r}$ via
\begin{equation}
    \mathbf{r}' = \mathbf{r} - \frac{\mathbf{e}}{s}.
\end{equation}
Finally we introduce the abbreviated notation $\alpha_i = \zeta_i,\hat{\nu}_i = \mathbf{r}_i,s_i,\lambda_i,\mathcal{I}_i$.
With this notation in place we can write the spin-down electron hopping matrix element in the cluster basis as
\begin{widetext}
\begin{multline} \label{eq:coefficients}
    \bra{\left\{\mathcal{I}'\right\},\mathscr{C}'}\left(-t\sum_{\mathbf{e}}c^\dagger_{\mathbf{x}+\mathbf{e},\downarrow} c_{\mathbf{x},\downarrow} \right) \ket{\left\{\mathcal{I}\right\},\mathscr{C}}
    = -t \bigoplus_{i=1}^{N'_c} \bigoplus_{j=1}^{N_c} \bra{0,\zeta'_i} C_{\alpha'_i} \sum_{\mathbf{e}} c^\dagger_{\mathbf{x}+\mathbf{e},\downarrow} c_{\mathbf{x},\downarrow} C^\dagger_{\alpha_j}\ket{0,\zeta_j} \\
    = -t \bigoplus_{i=1}^{N'_c} \bigoplus_{j=1}^{N_c} \bra{0,\zeta'_i}\!\! \sum_{\{\mathbf{x}_i\}\in\zeta'_i} \!\! c_{\mathbf{x}_{i,1}} \cdots c_{\mathbf{x}_{i,\nu'_i}} \Phi^*_{\alpha'_i}(\mathbf{x}_{i,1},\dots,\mathbf{x}_{i,\nu'_i}) \sum_{\mathbf{e}} c^\dagger_{\mathbf{x}+\mathbf{e},\downarrow} c_{\mathbf{x},\downarrow} \!\! \sum_{\{\mathbf{y}_j\}\in\zeta_j} \!\! \Phi_{\alpha_j}(\mathbf{y}_{j,1},\dots,\mathbf{y}_{j,\nu_j}) c^\dagger_{\mathbf{y}_{j,\nu_j}}\cdots c^\dagger_{\mathbf{y}_{j,1}}\ket{0,\zeta_j} \\
    = -t \, \delta_{\nu',\nu} \delta_{s',s} \delta_{\mathbf{r}',\mathbf{r}-\mathbf{e}/s} \sum_{\{\mathbf{x}\}\in Z'_c} \Phi^*_{\alpha'_1\dots\alpha'_{N'_c}}(\mathbf{x}_1,\dots,\mathbf{x}_{\nu'}) \Phi_{\alpha_1\dots\alpha_{N_c}}(\mathbf{x}_1,\dots,\mathbf{x}_\nu) \\
    \equiv -t\,g^{(N_c\to N'_c)}_{\alpha'_1\dots\alpha'_{N'_c}; \alpha_1\dots\alpha_{N_c}}.
\end{multline}
In the third line we define the multicluster, multielectron wave functions as the product of the multielectron wave functions in the clusters in question, 
\begin{equation} \label{eq:multiclusterWF}
    \Phi_{\alpha_1\dots\alpha_N}(\mathbf{x}_1,\dots,\mathbf{x}_\nu) = \Phi_{\alpha_1}(\mathbf{x}_1,\dots,\mathbf{x}_{\nu_1}) \cdots \Phi_{\alpha_N}(\mathbf{x}_{\nu-\nu_N+1},\dots,\mathbf{x}_{\nu_N}),
\end{equation}
and in the final line we define the symbol $g^{(N_c\to N'_c)}_{\alpha'_1\dots\alpha'_{N'_c};\alpha_1\dots\alpha_{N_c}}$ to represent the entire inner product of the initial and final multielectron wave functions, including the $\delta$-functions enforcing conservation of electron number and total cluster size, and the motion of the cluster center of mass. 
\end{widetext}

\section{Internal spectra of large clusters}
\label{sec:clusterspectra}
\begin{figure}
    \centering
    \includegraphics[width=\columnwidth]{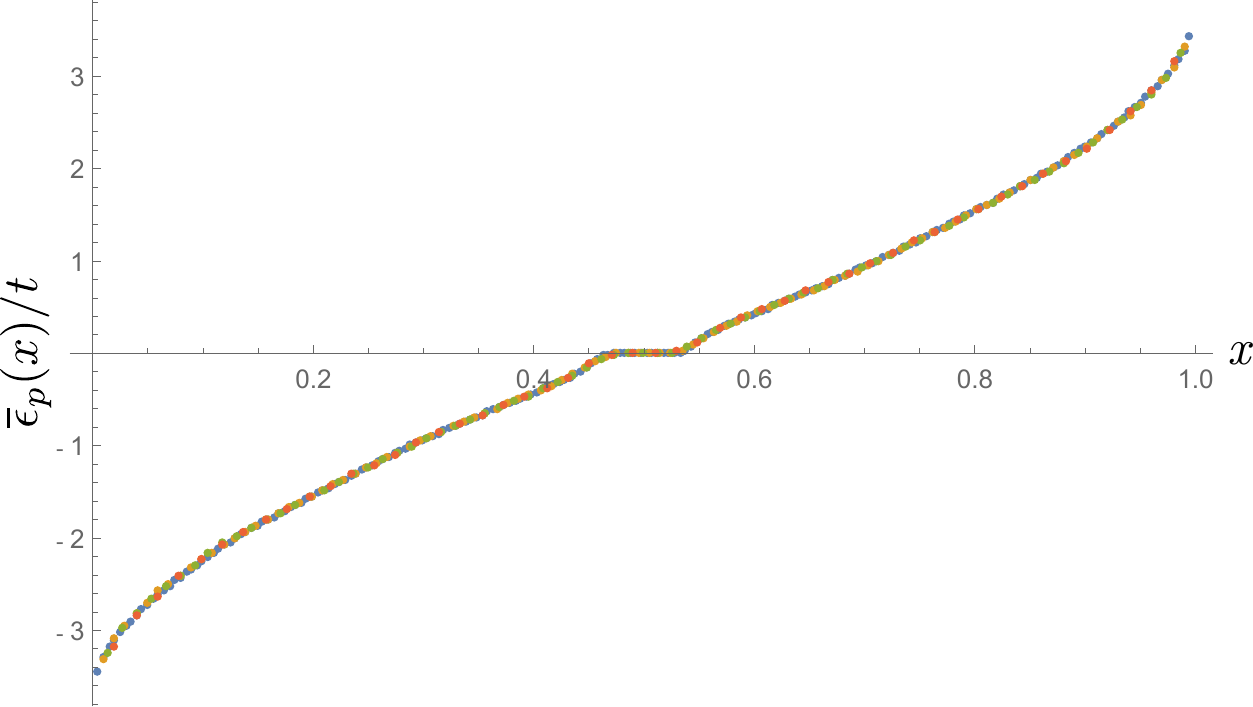}
    \caption{The average spectra for clusters of size 50 (red), 75 (green), 100 (yellow), and 200 (blue). 
    For each size we generate 100 random clusters, find and sort the energies of the $s$ eigenstates, then average each energy in order.
    The horizontal axis is the index of the state over $s+1$. 
    We see that they all follow the same curve which we call $\bar\epsilon_p(x)$, with the average energies for a cluster of size $s$ very closely reproduced as the values of $\bar\epsilon_p(x)$ at the points $x_i=i/(s+1)$ for $i=1,2,\dots,s$.}
    \label{fig:clusterspectra}
\end{figure}

\vspace{-12pt}
Here we investigate how well the single-particle spectra within individual large clusters are approximated by the average over cluster shapes.
We examine this question numerically: for a range of different sizes we generate a large number of random clusters, obtaining a representative sample of the vast number of cluster shapes, then diagonalize the nearest-neighbor hopping Hamiltonian and average the energies thus obtained.
The discrete indices labeling these averaged energies, $i = 1,2,\dots, s$ are then related to samplings of a continuous parameter $x$ running from 0 to 1 as $x_i = i/(s+1)$. 
The result of this procedure for $s = 50, 75, 100,$ and $200$, averaging over 100 different shapes each, is shown in \cref{fig:clusterspectra}. 
The average spectra for different sized clusters are found to trace the same curve as a function of the parameter $x$, and we define this function to be the continuous spectrum $\bar{\epsilon}(x)$.

\section{Evaluation of scaling Green's functions, $G_3$} \label{sec:G3details}
\vspace{-12pt}
The values of the exponents $\gamma_\pm$ and $\Delta_\pm$ and the behavior of the functions $\Omega_\pm(\mathbf{k})$ and $\theta_\pm(s)$ appearing in the scaling ansatz for the Green's function in the cubic regime, $G_{3,\pm}$ are determined by employing the spectral representation of the Green's function for complex frequency $z$,
\begin{equation}
    G_{3,\pm}(z,\mathbf{k},s) = \int_{-\infty}^\infty \dd\omega \frac{\rho_{3,\pm}(\omega,\mathbf{k},s)}{z-\omega}.
\end{equation}
Imposing the appropriate positivity constraints on the spectral density for bosons and fermions, $-\Im G_{3,-}(\omega+i0) > 0$ and $-\omega\Im G_{3,+}(\omega+i0)>0$ we immediately obtain
\begin{gather}
    -\pi\Delta_- < \theta_-(s) < \pi\Delta_- \\
    \pi\Delta_+ < \theta_+(s) < \pi(1-\Delta_+).
\end{gather}

Fourier transforming the spectral representation of $G_{3,\pm}$ to imaginary time $\tau$ we have
\begin{equation}
    G_{3,\pm}(\tau,\mathbf{k},s) = \begin{cases}
        -\int_0^\infty \dd\omega \rho_{3,\pm}(\omega,\mathbf{k},s) e^{-\omega\tau}, & \tau > 0\\
        \int_0^\infty \dd\omega \rho_{3,\pm}(-\omega,\mathbf{k},s) e^{-\omega\abs{\tau}}, & \tau < 0,
    \end{cases}
\end{equation}
with which we can write the imaginary time representation of the Green's functions,
\begin{multline}
    G_{3,\pm}(\tau,\mathbf{k},s) = -\sgn(\tau)\frac{s^{\gamma_\pm}\Gamma(2\Delta_\pm)}{\pi \Omega_\pm(\mathbf{k})^{2\Delta_\pm}\abs{\tau}^{2\Delta_\pm}} \\
    \times \sin\left(\pi\Delta_\pm + \sgn(\tau)\theta_\pm(s)\right).
\end{multline}
Substituting this into \cref{eq:Sigma+,eq:Sigma-}, the cubic terms of the self-energies, which we denote $\Sigma_{3,\pm}(\tau,\mathbf{k},s)$, can then be written explicitly. 

Alternatively, we can start from the Dyson equation, 
\begin{equation}
    G_{3,\pm}(z,\mathbf{k},s)^{-1} = z + \mu_\pm(s) - \Sigma_{3,\pm}(z,\mathbf{k},s).
\end{equation}
If we assume that the zero-frequency part of the self-energy cancels the chemical potential, then for $z\neq0$ we have
\begin{equation}
    \Sigma_{3,\pm}(z,\mathbf{k},s) = z - s^{-\gamma_\pm}\Omega_\pm(\mathbf{k})^{2\Delta_\pm} e^{i(\pi\Delta_\pm + \theta_\pm(s))}z^{1-2\Delta_\pm}.
\end{equation}
Using a spectral representation for $\Sigma$ in terms of a function $\sigma$ (as $G$ is written in terms of $\rho$) and a Fourier transform of this spectral representation we obtain a second expression for the self-energies in terms of the imaginary time $\tau$,
\begin{multline}
    \Sigma_{3,\pm}(\tau,\mathbf{k},s) = -\sgn(\tau)s^{-\gamma_\pm} \frac{\Gamma(2-2\Delta_\pm)}{\pi\abs{\tau}^{2-2\Delta_\pm}}\Omega_\pm(\mathbf{k})^{2\Delta_\pm} \\
    \times\sin\left(\pi\Delta_\pm + \sgn(\tau)\theta_\pm(s)\right).
\end{multline}
Equating the two representations we have thus obtained for $\Sigma_{3,\pm}(\tau,\mathbf{k},s)$ and demanding consistency for generic $s$ and in the limit $s\to s_\xi$ gives $\Delta_\pm = 1/3$ and $\gamma_\pm = -1/3$, and also constrains $\Omega_\pm(\mathbf{k})\propto t\,\phi(\mathbf{k})$.

\bibliography{references}

\end{document}